\documentclass[11pt]{article}

\usepackage[final]{acl}

\usepackage{times}
\usepackage{latexsym}
\usepackage{booktabs}
\usepackage{multirow}
\usepackage{CJKutf8}
\usepackage{amssymb}
\usepackage[T1]{fontenc}

\usepackage[utf8]{inputenc}

\usepackage{microtype}

\usepackage{inconsolata}

\usepackage{graphicx}
\usepackage{array}    
\usepackage{xspace}
\usepackage{amsmath}
\usepackage{url}
\newcommand{\asyembedder}{CARE\xspace}

\title{Benchmarking and Enabling Efficient Chinese Medical Retrieval via Asymmetric Encoders}

\author{
  \textbf{Angqing Jiang}\textsuperscript{1,2,3} \quad
  \textbf{Jianlyu Chen}\textsuperscript{1,2} \quad
  \textbf{Zhe Fang}\textsuperscript{2,3} \quad
  \textbf{Yongcan Wang}\textsuperscript{2,3} \quad
  \textbf{Xinpeng Li}\textsuperscript{2,3} \\
  \textbf{Keyu Ding}\textsuperscript{4}\thanks{Corresponding author} \quad
  \textbf{Defu Lian}\textsuperscript{1,2} \\
  \textsuperscript{1}University of Science and Technology of China \\
  \textsuperscript{2}State Key Laboratory of Cognitive Intelligence \quad
  \textsuperscript{3}iFlytek Research \\
  \textsuperscript{4}HeFei Institute of Technology \\
  \texttt{\{philipgaq, chenjianlv\}@mail.ustc.edu.cn} \quad
  \texttt{kyding@hfit.edu.cn} \\
  \texttt{liandefu@ustc.edu.cn} \quad
  \texttt{\{zhefang, ycwang12, xpli\}@iflytek.com}
}

\begin{document}
\maketitle
\begin{abstract}
Effective medical text retrieval requires both high accuracy and low latency. While LLM-based embedding models possess powerful retrieval capabilities, their prohibitive latency and high computational cost limit their application in real-time scenarios. 
Furthermore, the lack of comprehensive and high-fidelity benchmarks hinders progress in Chinese medical text retrieval. In this work, we introduce the \textbf{C}hinese \textbf{Med}ical \textbf{T}ext \textbf{E}mbedding \textbf{B}enchmark (\textbf{CMedTEB}), a benchmark spanning three kinds of practical embedding tasks: retrieval, reranking, and semantic textual similarity (STS). Distinct from purely automated datasets, CMedTEB is curated via a rigorous {multi-LLM voting pipeline validated by clinical experts}, ensuring gold-standard label quality while effectively mitigating annotation noise.
On this foundation, we propose the \textbf{C}hinese Medical \textbf{A}symmetric \textbf{RE}triever (\textbf{CARE}), an asymmetric architecture that pairs a lightweight BERT-style encoder for online query encoding with a powerful LLM-based encoder for offline document encoding. However, optimizing such an asymmetric retriever with two structurally different encoders presents distinctive challenges. To address this, we introduce a novel two-stage training strategy that progressively bridges the query and document representations. Extensive experiments demonstrate that CARE surpasses state-of-the-art symmetric models on CMedTEB, achieving superior retrieval performance without increasing inference latency.
\end{abstract}

\section{Introduction}

\begin{figure}[t]
  \centering
  \includegraphics[width=\columnwidth]{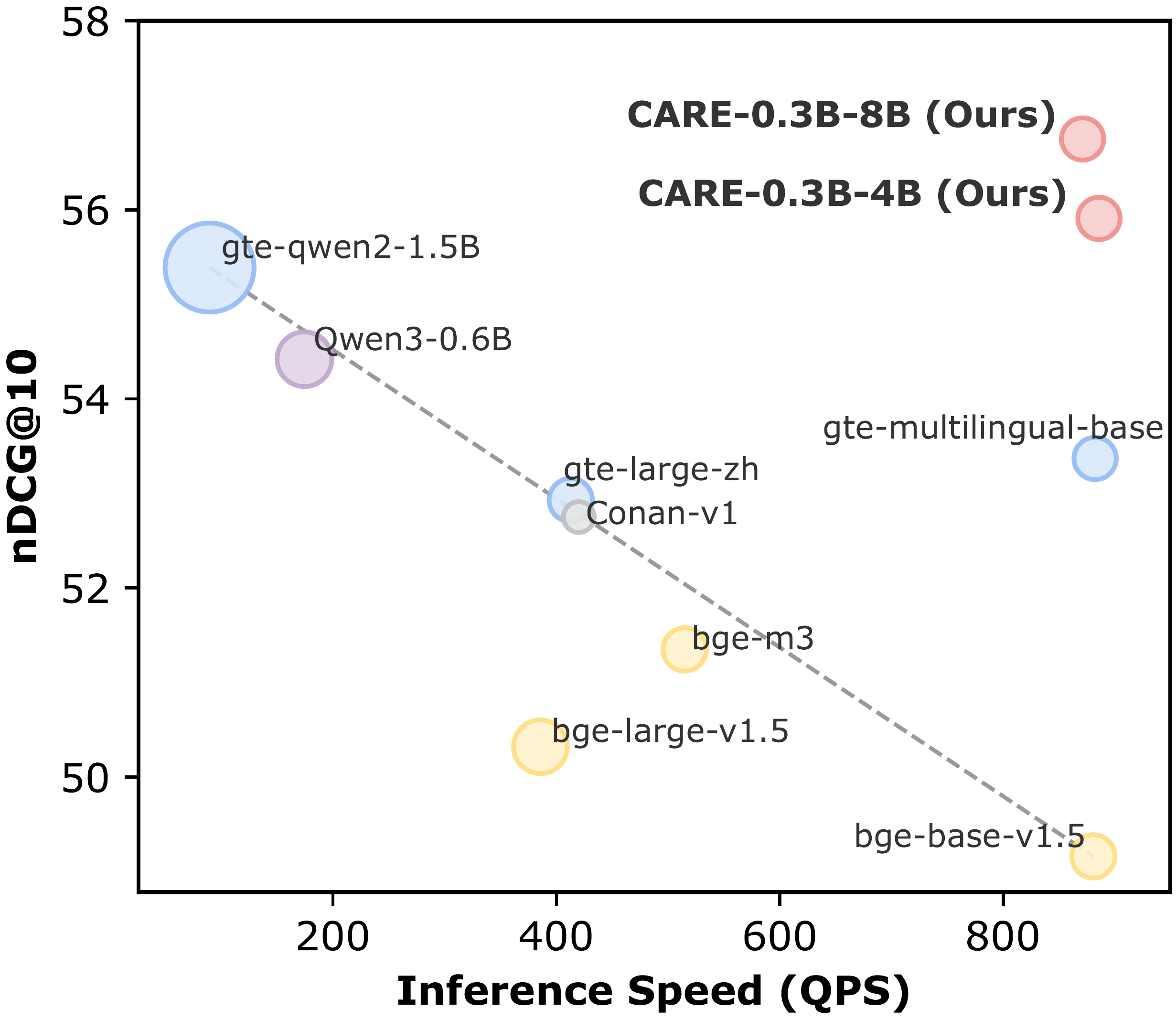}
  \caption{\textbf{Efficiency-performance trade-off on CMedTEB Retrieval.} The x-axis shows queries per second (QPS) on a single A100 80GB GPU, while the y-axis reports nDCG@10. Notably, \textbf{\asyembedder breaks the conventional trade-off}: it matches the high retrieval quality of heavy LLM-based models while sustaining the high throughput of lightweight BERT-style models.}
  \vspace{-6pt}
  \label{fig:latency_new}
\end{figure}

Text embedding models are essential for a wide range of natural language processing (NLP) tasks, including retrieval, reranking, and classification~\citep{reimers2019sentence}. Their role is crucial in retrieval-augmented generation (RAG) systems \citep{lewis2020retrieval, zhang2026stable}, which leverage external knowledge to enhance large language models (LLMs). In specialized domains such as healthcare, where LLMs often lack deep expert knowledge, accurate and low-latency access to medical knowledge can enhance clinical decision support and mitigate hallucinations in RAG, making domain-specific, low-latency embeddings indispensable.


Despite recent rapid progress in general-domain embedding models (e.g., BGE~\citep{chen2024bge}, GTE~\citep{li2023towards}, Qwen3-Embedding~\citep{zhang2025qwen3}), Chinese medical text embedding has received limited attention. 
Existing benchmarks like C-MTEB~\citep{xiao2024c} include only two Chinese medical retrieval datasets, both suffer from annotation sparsity and substantial false negatives (see analysis in Section~\ref{sec:medteb}). While recent benchmark CMIRB~\citep{li-etal-2025-automir} scale up data via single-LLM generation (e.g., ChatGPT), it lacks rigorous multi-model consensus and expert verification. Besides, it focuses exclusively on retrieval tasks while still incorporating legacy datasets with false negative issues, leaving a gap for a comprehensive, high-fidelity benchmark.
Moreover, while current state-of-the-art embedding models are mostly LLM-based~\citep{lee2024nv, lin2025causal2vec}, they incur prohibitive latency and high computational costs, which limit their applications in latency-sensitive scenarios such as real-time medical QA. This highlights the challenge of accuracy-latency trade-off.


To address the challenges of false negative noise and ensure robust evaluation beyond single LLM generated labels, we introduce the \textbf{C}hinese \textbf{Med}ical \textbf{T}ext \textbf{E}mbedding \textbf{B}enchmark (\textbf{CMedTEB}). CMedTEB consists of three newly curated tasks: retrieval, reranking, and medical synonym STS. To ensure high data quality, we employ a \textbf{multi-LLM consensus pipeline} for annotation. Our experiments indicate that even advanced general-purpose embedders underperform on CMedTEB, highlighting the benchmark's difficulty and value for domain-specific optimization.


To address the trade-off between retrieval accuracy and online inference latency in medical retrieval, we propose \textbf{C}hinese Medical \textbf{A}symmetric \textbf{RE}triever (\textbf{CARE}), an asymmetric architecture that pairs a lightweight BERT-style encoder for online query encoding with a powerful LLM-based encoder for offline document encoding. By employing a novel two-stage alignment strategy, CARE effectively bridges the semantic gap, achieving LLM-level accuracy with BERT-level online latency. As shown in Figure~\ref{fig:latency_new}, while most embedding models exhibit a clear accuracy-latency trade-off, \asyembedder breaks this trend. It matches the retrieval accuracy of large-scale LLM-based embedding models while sustaining Queries Per Second (QPS) levels comparable to small-size BERT-style embedding models. We further observe that as the document encoder scales up, the asymmetric model progressively closes the gap with LLM-based embedding models, offering a practical path to scale retrieval performance without sacrificing online latency.


The primary contributions of our work are summarized as follows:

$\bullet$~We introduce \textbf{CMedTEB}, a comprehensive, high-fidelity benchmark for Chinese medical text embedding, establishing a reliable standard for medical domain-specific evaluation.

$\bullet$~We propose the \textbf{CARE framework}, which integrates an asymmetric inference architecture with a progressive two-stage training strategy. Our design reconciles the efficiency-accuracy conflict, enabling LLM-level retrieval quality with BERT-style inference latency.

$\bullet$~To facilitate future research in medical text retrieval, we will open-source our benchmark, models, and code at our repository\footnote{\url{https://github.com/PhilipGAQ/CARE}}.
\section{Related Work}

\paragraph{Medical Retrieval Benchmarks.}
MTEB~\citep{muennighoff2022mteb} provides a comprehensive benchmark across languages and tasks, and its Chinese extension C-MTEB~\citep{xiao2024c} includes several Chinese embedding model datasets. 
However, domain-specific evaluation of Chinese medical remains scarce. 
Existing Chinese medical benchmarks in C-MTEB, such as CmedqaRetrieval~\citep{zhang2017chinese} and MedicalRetrieval~\citep{Long2022MultiCPRAM}, both exhibit annotation noise and false negatives (see analysis in Appendix~\ref{app:open_source_analysis}). Although recent works like CMIRB~\citep{li-etal-2025-automir} attempt to address data scarcity via automated generation, they rely on unverified single-source LLM judgments and incorporate noisy legacy datasets. Furthermore, it focuses exclusively to retrieval tasks. As a result, the field still lacks a comprehensive, high‑quality, and human-verified benchmark for Chinese medical text embedding.

\paragraph{Embedding Models.}
Text embedding models have advanced rapidly alongside pretrained language models. Early works such as Contriever~\citep{izacard2021unsupervised} explored unsupervised contrastive pretraining, while more recent models like E5~\citep{wang2022text}, GTE~\citep{li2023towards}, and the BGE series~\citep{chen2024bge} leveraged large-scale contrastive pretraining to obtain strong general-purpose embeddings. In the biomedical domain, specialized models such as MedCPT~\citep{10.1093/bioinformatics/btad651} and BMRetriever~\citep{xu-etal-2024-bmretriever} leverage large-scale medical corpus and tuning language models for enhanced retrieval. Recently, decoder-only embedding models such as Qwen3-Embedding~\citep{zhang2025qwen3}, bge-en-icl~\citep{li2024making}, and NV-Embed~\citep{lee2024nv} have achieved state-of-the-art performance on MTEB~\citep{muennighoff2022mteb}.

Despite these advances, most LLM-based models contain billions of parameters. While they deliver strong accuracy, their high latency and computational overhead make them impractical for latency-sensitive applications such as real-time medical retrieval. This gap highlights the urgent need for lightweight yet effective embedding models in specialized domains.

\paragraph{Asymmetric Retrieval Architecture.}

While existing works of asymmetric architectures offer a promising path for retrieval efficiency, achieving effective alignment between disparate encoders remains a significant challenge. Existing approaches primarily follow two paradigms: (1) Homogeneous Distillation: Methods like KALE~\citep{wang2023query,campos2023quick} prune layers from a teacher model to initialize a student. However, this creates a rigid architectural dependency, constraining the student model to the specific design of the teacher. (2) Heterogeneous Alignment: Recent works such as ScalingNote~\citep{huang2024scalingnote} or HotelMatch~\citep{askari2025hotelmatch} aligns different architectures. However, aligning heterogeneous encoders presents a significant semantic gap. Without tailored training, lightweight models fail to adapt to the complex embedding space of large document encoders, resulting in suboptimal retrieval performance.

In contrast, we propose a novel \textbf{two-stage training strategy} that bridges the representational gap between the lightweight query encoder and the LLM-based document encoder. This progressive approach ensures stable convergence, enabling the smaller model to accurately map inputs into the rich semantic space of the larger model without architectural constraints.
\section{CMedTEB}
\label{sec:medteb}

\begin{figure*}[t]
    \centering
    \includegraphics[width=0.95\linewidth]{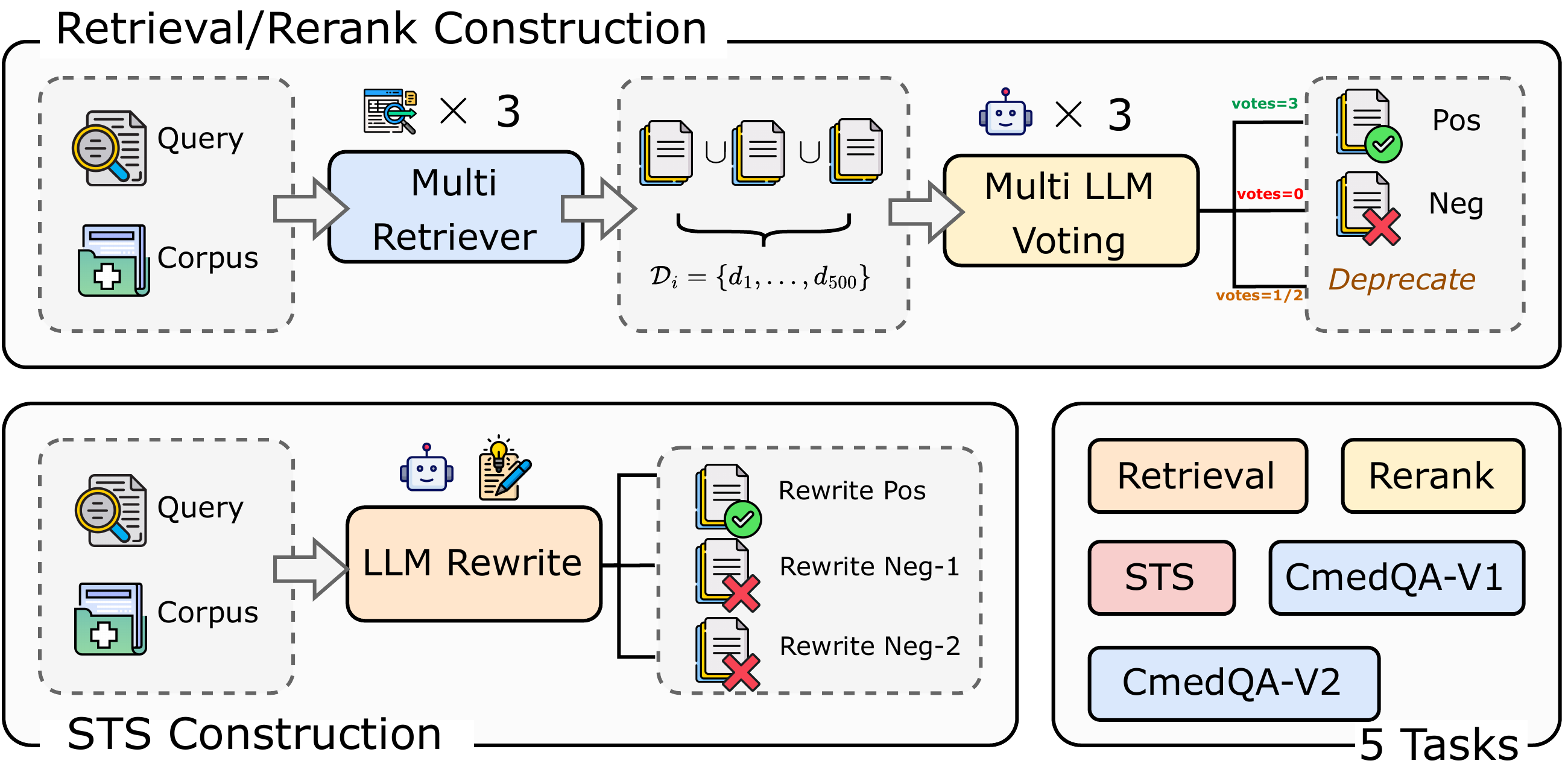}
    \caption{\textbf{Workflow for constructing the CMedTEB benchmark.} The Figure shows the distinct curation strategies for discriminative tasks (Retrieval/Rerank) and semantic similarity tasks (STS).}
    \label{fig:pipeline_overview}
\end{figure*}

Chinese medical text embedding benchmarks remain scarce. Among the few available benchmarks, CmedqaRetrieval~\citep{zhang2017chinese} and MedicalRetrieval~\citep{Long2022MultiCPRAM} are well known and widely used. These datasets are constructed primarily from human-labeled query-answer pairs sourced from online medical Q\&A platforms, such as patient inquiries and physician responses. However, this methodology inherently ignores potentially relevant yet unlabeled candidate answers associated with other pairs. The medical domain further exhibits \textit{topic intensity}: common diseases or medications often generate a large volume of semantically similar queries and answers, increasing the risk of false negatives (See Table~\ref{tab:false_neg_medretri} for examples).


To quantify this issue, we performed an analysis of current benchmarks using LLM-assisted annotation followed by {human verification} (details in Appendix~\ref{app:open_source_analysis}). The results suggested a significant presence of \textbf{potential false negatives}: on average, each query is associated with approximately \textbf{9 False Negatives} in MedicalRetrieval, and \textbf{19 False Negatives} in CmedqaRetrieval. To validate the reliability of these findings, we employed human assessors to re-judge a random sample of 500 pairs which LLM labeled as false negatives. yielding a \textbf{92\% consistency rate} with model's suggestions, providing strong evidence that existing retrieval benchmarks suffer from notable annotation noises.

To address the challenges of false negatives and ensure robust evaluation beyond single LLM generated labels, we construct CMedTEB (Figure~\ref{fig:pipeline_overview}) via a rigorous multi-LLM consensus pipeline. CMedTEB comprises three new tasks: Retrieval, Reranking, and Synonym STS, along with two high-quality, human-verified existing public datasets CMedQAv1-reranking~\citep{zhang2017chinese} and CMedQAv2-reranking~\citep{8548603}. We construct the document corpus using data from \textbf{XunYiWenYao}\footnote{\url{https://www.xywy.com/}}, and sampling queries from logs of an online medical service (construction details and anonymization steps on queries and documents are available in Appendix~\ref{app:medteb_detail}).  

\subsection{Construction Method}

\paragraph{Retrieval}
Prior studies like AIR-Bench~\citep{chen2024air} and \citet{thomas2024large} demonstrate the reliability of LLM-generated relevance labels in information retrieval benchmark. Building on these findings, we adopt a multi-LLM labeling pipeline.
Given a query $q_i \in \mathcal{Q}$, we used gte-multilingual-base, bge-m3, Conan-embedding-v1 to retrieve and gather a candidate pool of top-500 documents $\mathcal{D}_i = \{d_1, \dots, d_{500}\}$. Three strong LLMs, DeepSeek-V3~\citep{liu2024deepseek}, Doubao-1.5-Pro~\citep{guo2025seed1} and GPT-4o~\citep{hurst2024gpt} then rated each $(q_i, d_j)$ pair on a 5-point relevance scale. To ensure label quality, a document was retained as positive only when all three LLMs agreed, while pairs with partial agreement (only 1 or 2 agreements) were discarded. The final retrieval dataset comprise a query set $\mathcal{Q}$, a refined corpus $\mathcal{D}' \subseteq \mathcal{D}$, and relevance labels $\mathcal{R} = \{(q_i, d_j, y_{ij}) \mid y_{ij} \in \{0,1\}\}$.

\paragraph{Rerank}
For each query $q_i \in \mathcal{Q}$, we derive positives $P_i=\{d_j\in\mathcal{D}': y_{ij}=1\}$ and negatives $N_i=\{d_j\in\mathcal{D}': y_{ij}=0\}$ labeled from the same multi-LLM labeling pipeline as in Retrieval. The reranking dataset is a collection of triplets $\mathcal{T}_{\textbf{Rerank}}=\{(q_i,\mathcal{P}_i,\mathcal{N}_i)\}$, where $\mathcal{P}_i$ is a list sampled from $P_i$ and $\mathcal{N}_i$ is a list sampled from $N_i$.

\paragraph{STS}
We first build a medical synonym dictionary with domain experts. For each $q_i \in \mathcal{Q}$, GPT-4o generates three sentences: a positive $s^+_i$ (synonym substitution with semantics preserved), a hard negative $s^-_{i,1}$ (synonym substitution with semantics changed), and an easy negative $s^-_{i,2}$ (no synonym substitution with semantics changed). We then sample $ s_i \in \{s_i^{+}, s_{i,1}^{-}, s_{i,2}^{-}\} $ and pair it with $ q_i $ to form $ (q_i, s_i, y_i) $, where $ y_i=\mathbf{1}[\,s_i=s_i^{+}\,]\in\{0,1\} $. The dataset is $ \mathcal{T}_{\text{STS}}=\{(q_i,s_i,y_i)\} $, evaluating fine-grained synonym understanding.

\begin{table}[t]
\centering
\small
\begin{tabular}{lrrc}
\toprule
\textbf{Task} & \textbf{Test} & \textbf{Train} & \textbf{Main Metric} \\
\midrule
\multicolumn{4}{l}{\textit{New tasks}} \\
\midrule
Retrieval      & 734   & \multirow{2}{*}{20,000} & nDCG@10 \\
Rerank      & 1,128 &                         & MAP@10     \\
Synonym STS    & 5,000 & 10,000                  & Pearson \\
\midrule
\multicolumn{4}{l}{\textit{Public datasets}} \\
\midrule
CMedQA-v1-rk.  & 1,000 & \multirow{2}{*}{50,000} & MAP@10     \\
CMedQA-v2-rk.  & 1,000 &                         & MAP@10     \\
\bottomrule
\end{tabular}
\caption{CMedTEB statistics}
\label{tab:medteb_stats}
\vspace{-6pt}

\end{table}

\subsection{Quality Analysis}
\label{subsec:eval_medteb}

The statistics of CMedTEB are summarized in Table~\ref{tab:medteb_stats}, with detailed breakdowns provided in Appendix~\ref{app:medteb_stats_app}.

\paragraph{Annotation Accuracy Assessment.}
To ensure annotation reliability, we computed Fleiss' Kappa~\citep{fleiss1971measuring} among the results of three LLMs in our pipeline, yielding a score of \textbf{0.731}, indicating substantial agreement. Furthermore, a clinical expert independently re-annotated a large-scale sample of 5,000 query-document pairs, achieving a \textbf{93.3\%} agreement rate with our final labels. These metrics confirm that our automated pipeline aligns closely with clinical expertise.

\paragraph{Task Difficulty and Distinctness.}
To demonstrate the necessity of this benchmark, we evaluated existing general-domain embedding models (full zero-shot results are detailed in Appendix Table~\ref{tab:zero_shot_medteb}). We observe a sharp performance contrast: while models achieve high accuracy on legacy CMedQA tasks (Avg.~85.15), their performance drops drastically on CMedTEB new tasks (Avg.~57.85). Moreover, the Spearman rank correlation between model rankings on CMedQA versus CMedTEB new tasks is notably weak ($\rho=0.354, p=0.215 \gg 0.05$). This statistical lack of significant correlation indicates that CMedTEB is \textbf{non-redundant}, aiming to evaluate medical semantic capabilities overlooked by prior datasets. Finally, while massive decoder-only models like Qwen3-Embedding-8B achieve stronger results (Avg.~64.52), their prohibitive latency underscores the urgent need for efficient, domain-specialized solutions.
\section{CARE}
\label{sec:method}

\begin{figure*}[t]
    \centering
    \includegraphics[width=0.95\linewidth]{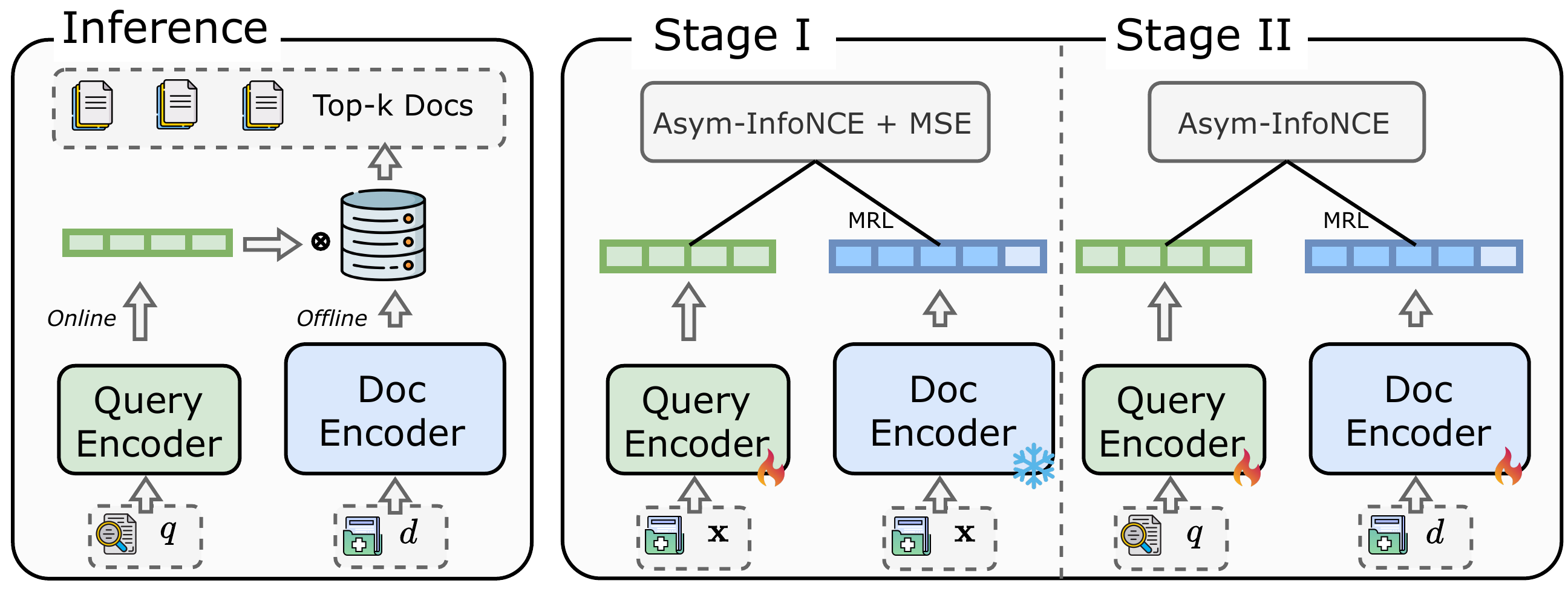}
    \caption{\textbf{Inference and Training pipeline for asymmetric embedding model.} Stage \uppercase\expandafter{\romannumeral1}: Query encoder is trained to align with the frozen document encoder using Asym-InfoNCE and MSE losses. Stage \uppercase\expandafter{\romannumeral2}: Both encoders are jointly fine-tuned with Asym-InfoNCE loss on retrieval data.}
    \label{fig:model_overview}
    \vspace{-8pt}
\end{figure*}


To bridge the gap between retrieval accuracy and inference latency observed in CMedTEB, Section~\ref{subsec:eval_medteb}, we propose the \textbf{Chinese Medical Asymmetric REtriever (CARE)}. As illustrated in Figure~\ref{fig:model_overview}, CARE adopts a decoupled architecture where a powerful LLM serves as the offline document encoder to capture rich semantics, while a lightweight BERT model acts as the online query encoder to ensure low-latency inference.

In this section, we describe our high-quality training data construction for medical domain, and a two-stage training strategy designed for our asymmetric embedding architecture.


\subsection{Training Data Construction}
\label{subsec:data_construction}

Standard hard negative mining fails in the medical domain due to \textit{topic intensity}, where abundant latent positives lead to severe false negatives. We address this via a \textbf{diversity-aware curation pipeline} applied independently to query and document sets. The process initializes a vector index with 5,000 seed samples. For each new candidate \(x\), we retrieve its top-\(k\) neighbors from the evolving index; if the count of neighbors exceeding similarity threshold \(t\) surpasses \(n\), \(x\) is discarded as redundant. Otherwise, it is added to update the distribution. Finally, we employ GPT-4o to verify top-50 retrieved candidates, distinguishing hard negatives from false positives. This yields \textbf{500K high-fidelity triples} $(q, d^+, d^-)$. Details are presented in Appendix~\ref{app:data_diversify}.

\subsection{Asymmetric Embedding Architecture}
While LLM-based embedders achieve state-of-the-art retrieval accuracy, their high computational cost and latency are prohibitive for real-time applications. To resolve this trade-off, we propose an \textbf{Asymmetric embedding architecture}, which pairs a lightweight query encoder $E_Q$ for fast online inference with a powerful document encoder $E_D$ whose embeddings are pre-computed offline.

To establish strong foundation before alignment, we first initalize $E_Q$ and $E_D$ independently with contrastive learning. Crucially, $E_D$ employs {Matryoshka Representation Learning (MRL)}~\citep{kusupati2022matryoshka} to truncate its native embedding dimension to match the smaller $E_Q$. 

Despite dimensional compatibility, a significant {semantic gap} persists between the embedding spaces of these heterogeneous models. We bridge this gap via a progressive two-stage strategy: (1) \textbf{Query Encoder Alignment}, which maps the query encoder's space to the frozen document encoder; (2) \textbf{Joint Fine-Tuning}, which optimizes both encoders for end-to-end retrieval performance.

\subsubsection{Asymmetric Stage \uppercase\expandafter{\romannumeral1}: Query Encoder Alignment}
\label{subsec:stage1}

In this stage, we freeze the document encoder (the \textit{teacher}) and update only the query encoder (the \textit{student}) to align the student's space to the teacher's.

To bridge the representational gap without consuming an amount of labeled data, we propose a \textbf{Self-Contrastive} strategy over abundant unlabeled corpora. Specifically, we treat each input text $\mathbf{x}$ as a positive anchor for itself. This unsupervised paradigm effectively leverages millions of raw texts, allowing the alignment process to scale up free from the constraints of annotation. 

\paragraph{Objective Function}
We employ a hybrid objective to enforce alignment from two perspectives:

\textit{ Asymmetric Contrastive Loss.}
We use Asym-InfoNCE with the frozen document encoder as the teacher:
\begin{equation}
    \mathcal{L}^{Asym}_{\text{InfoNCE}} = -\log \frac{\exp(s^+ / \tau)}{\exp(s^+ / \tau) + \sum_{i=1}^{N} \exp(s^-_i / \tau)},
    \label{eq:infonce_asym}
\end{equation}
where $s^+ = \text{sim}(E_Q(\mathbf{x}), E_D(\mathbf{x}))$ represents the similarity of the same text's embedding across two encoders, and $s^-_i = \text{sim}(E_Q(\mathbf{x}), E_D(\mathbf{x}^-_i))$ represents similarity with in-batch negatives. This explicitly trains the student $E_Q$ to identify the teacher $E_D$'s representation of the same input against negative distractors.

\textit{MSE Loss.}
For stricter alignment, we minimize the L2 distance:
\begin{equation}
    \mathcal{L}_{\text{MSE}} = \|E_Q(\mathbf{x}) - E_D(\mathbf{x})\|^2_2,
\end{equation}
where embeddings are normalized. This penalizes absolute deviations in the embedding space.

The final objective is 
\begin{equation}
    \mathcal{L}_{\text{Stage 1}} = \lambda_1 \mathcal{L}^{Asym}_{\text{InfoNCE}} + \lambda_2 \mathcal{L}_{\text{MSE}}   
\end{equation}
We empirically set $\lambda_1=\lambda_2=1$ to balance soft ranking alignment (InfoNCE) with hard structural alignment (MSE).

\subsubsection{Asymmetric Stage \uppercase\expandafter{\romannumeral2}: Joint Fine-tuning}
\label{subsec:stage2}

After alignment, we unfreeze both encoders and perform end-to-end joint fine-tuning. The goal of this stage is to further enhance retrieval performance by jointly optimizing the two encoders to better discriminate between positive and negative documents. We adopt the \textbf{Asym-InfoNCE loss} as the sole objective:

\begin{equation}
    \mathcal{L}_{\text{Stage 2}} = -\log \frac{e^{s(q, d^+) / \tau}}{e^{s(q, d^+) / \tau} + \sum_{d^- \in \mathcal{N}} e^{s(q, d^-) / \tau}},
\end{equation}
where $s(q, d) = \text{sim}(E_Q(q), E_D(d))$, and $\mathcal{N}$ includes both in-batch negatives and the hard negatives. Unlike Stage 1, $q$ and $d^+$ are semantically relevant pairs. This end-to-end optimization directly refines the decision boundary for retrieval.
\section{Experiments}
\label{sec:experiments}

In our experiments, we aim to answer the following research questions:
\begin{itemize}
    \item \textbf{RQ1:} How does CARE compare with state-of-the-art baseline embedding models on CMedTEB?
    \item \textbf{RQ2:} {Can CARE match the performance of large symmetric models with \textbf{lightweight} inference?}
    \item \textbf{RQ3:} Is the proposed CARE framework superior to other efficient retrieval methods?
    \item \textbf{RQ4:} What are the contributions of different components in our training strategy?
\end{itemize}

\subsection{Setup}
\label{subsec:setup}
\paragraph{Training Data.}
We use the curated dataset (Section~\ref{subsec:data_construction}) and CMedTEB training splits. Although some baselines have previously seen CMedQA during pre-training, we explicitly include it to prevent potential 
performance degradation on this task.
\paragraph{Baselines.}
As we target efficient online deployment, we prioritize comparisons with \textit{lightweight} yet strong open-source models (e.g., BGE~\citep{xiao2024c}, GTE~\citep{li2023towards}, Conan~\citep{li2024conan}, Stella~\citep{zhang2024jasper}) and moderate-sized LLM embedders (Qwen3-Embedding~\citep{zhang2025qwen3}, gte-Qwen2~\citep{li2023towards}), rather than prohibitive large size models.
\paragraph{Implementation Details.}
The query encoder (\textbf{Med-Emb-base}) is initialized from \texttt{gte-multilingual-base}. Document encoders (\textbf{Med-Emb-4B/8B}) are fine-tuned from Qwen3 using LoRA. Implementation details are in Appendix~\ref{app:independent_initialization}. We apply Matryoshka Representation Learning (MRL) to project their native dimensions (2560/4096) to the query encoder's 768-dim space. We evaluate two asymmetric variants: \textbf{CARE-0.3B-4B} and \textbf{CARE-0.3B-8B}. Experiments run on 32$\times$A100 GPUs. Details are in Appendix~\ref{app:hyperparams}.

\subsection{Main results on CMedTEB (RQ1)}

\begin{table*}[t]
\centering
\small
\setlength{\tabcolsep}{5pt}
\resizebox{0.9\textwidth}{!}{ 
\begin{tabular}{lcccccccc}
\toprule
\multirow{2}{*}{\textbf{Model}} & \multirow{2}{*}{\textbf{Params (Q/D)}} & \textbf{CMed v1} & \textbf{CMed v2} & \textbf{Retrieval} & \textbf{Rerank} & \textbf{STS} & \multirow{2}{*}{\textbf{Avg}} \\
 & & MAP@10 & MAP@10 & nDCG@10 & MAP@10 & Pearson & \\
\midrule
\textit{Baselines} \\
\midrule
bge-small-zh-v1.5       & 24M / 24M     & 80.21 & 81.69 & 44.33 & 62.30 & 70.50 & 67.81 \\
bge-base-zh-v1.5        & 102M / 102M   & 83.37 & 83.31 & 49.16 & 66.73 & 76.24 & 71.76 \\
bge-large-zh-v1.5       & 326M / 326M   & 83.23 & 85.15 & 50.32 & 67.55 & 78.95 & 73.04 \\
bge-m3                  & 568M / 568M   & 82.98 & 83.32 & 51.35 & 66.90 & 78.34 & 72.58 \\
Conan-embedding-v1      & 326M / 326M   & \textbf{89.89} & \underline{88.77} & 52.75 & 69.31 & 81.49 & 76.44 \\
stella-base-zh-v3-1792d & 102M / 102M   & 87.16 & 88.28 & 53.31 & 69.56 & 80.52 & 75.77 \\
gte-multilingual-base   & 305M / 305M   & 86.21 & 86.37 & 53.37 & 69.38 & 82.36 & 75.54 \\
gte-base-zh             & 102M / 102M   & 85.31 & 86.44 & 52.62 & 69.35 & 79.73 & 74.69 \\
gte-large-zh            & 326M / 326M   & 85.44 & 86.97 & 52.93 & 69.97 & 81.48 & 75.36 \\
gte-Qwen2-1.5B-instruct & 1.78B / 1.78B & {87.68} & 87.15 & 55.39 & 72.35 & 85.50 & 77.61 \\
Qwen3-Embedding-0.6B    & 596M / 596M   & 85.58 & 86.09 & 54.42 & 70.94 & 80.42 & 75.49 \\
\midrule
\textit{Ours} \\
\midrule
CARE-0.3B-4B$\dagger$ & 305M / 4.02B  & 86.04 & 87.31 & \underline{55.91} & \underline{72.84} & \textbf{88.53} & \underline{78.13} \\
CARE-0.3B-8B$\dagger$ & 305M / 8.19B  & \underline{88.34} & \textbf{88.86} &\textbf{ 56.75} & \textbf{73.67} & \underline{87.07} & \textbf{78.94} \\
\bottomrule
\end{tabular}
}
\caption{
\textbf{Results of our models compare to the baselines on CMedTEB.}
Best results in \textbf{bold}, second-best in \underline{underline}.
Asymmetric models are marked with $\dagger$.
}
\label{tab:medteb_res}
\end{table*}

Table~\ref{tab:medteb_res} presents the evaluation of CARE series on the CMedTEB benchmark, alongside strong open-source baselines.
We observe two key findings:
(1) CARE establishes a new state of the art: the 0.3B$\text{-}$4B variant achieves an average score of 78.13, and the 0.3B$\text{-}$8B variant reaches 78.94, surpassing the strongest baseline gte-Qwen2-1.5B-instruct (77.61, a decoder-only model), despite using a much smaller query encoder.
(2) Both baseline models and our asymmetric variants exhibit consistent performance scaling with model size: enlarging the document encoder from 4B to 8B improves the average score by 0.81 with \textit{zero} increase in online query cost, verifying the superior accuracy-latency trade-off.

\subsection{Asymmetric vs. Symmetric Architectures (RQ2)}
\label{sec:asym_effect}

\begin{figure}[t]
    \centering
    \includegraphics[width=\columnwidth]{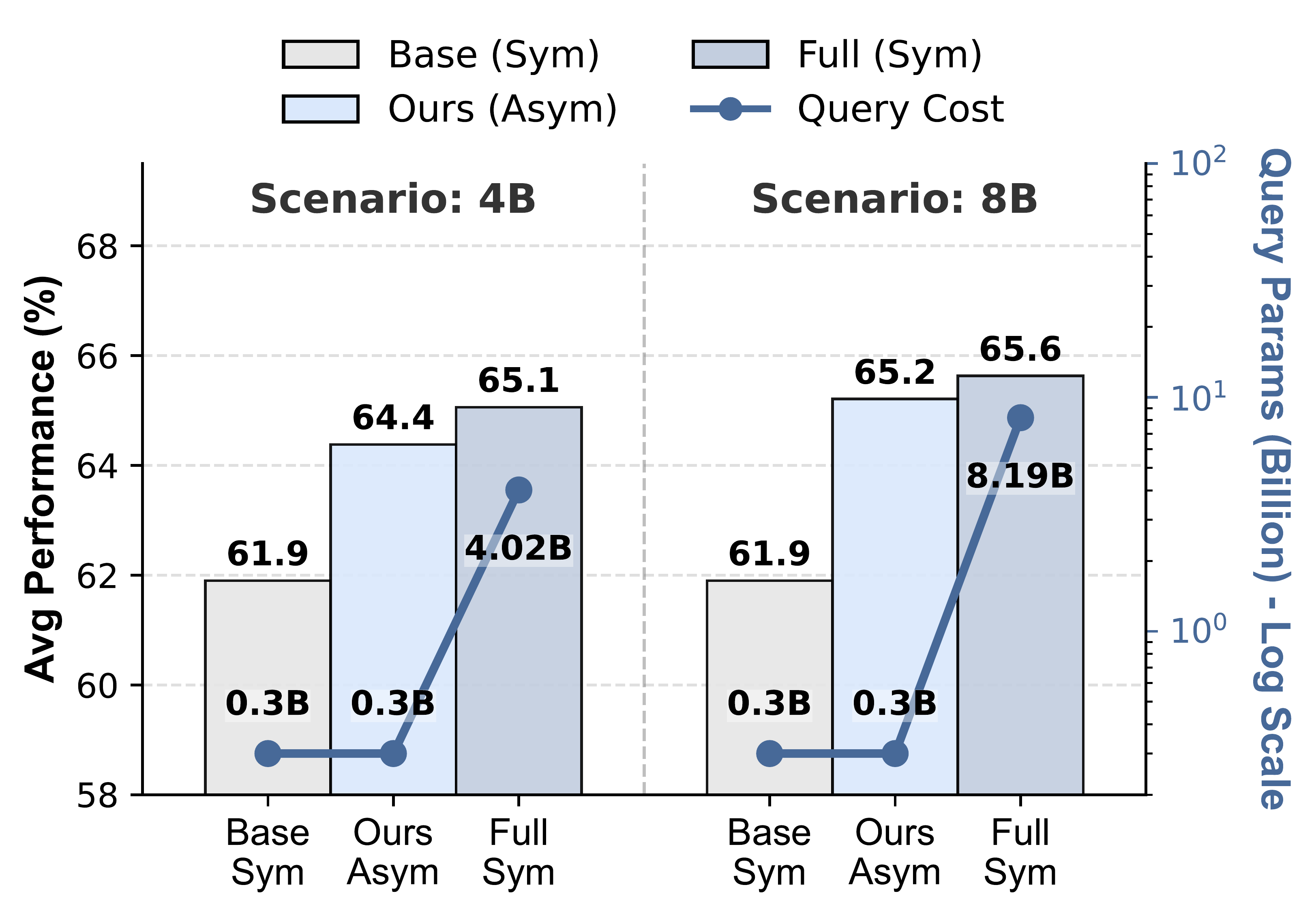} 
    \caption{\textbf{Asymmetric vs. Symmetric Architectures on performance and parameters.} We evaluate models in two scenarios: leveraging power from 4B (left) and 8B (right) document encoders. The bars denote average performance on CMedTEB Retrieval and Rerank (higher is better), while the blue line represents the inference cost in terms of Query Encoder parameters (log scale, lower is better). CARE (Ours) bridge the performance gap to symmetric giants (Full) while maintaining same online inference cost to baselines.}
    \label{fig:performance_efficiency}
\end{figure}

Figure~\ref{fig:performance_efficiency} visualizes the performance trade-off between symmetric (Med-Emb) and asymmetric (CARE) architectures. While symmetric models (Med-Emb-4B/8B) set a high performance ceiling, they incur prohibitive computational costs, indicated by the sharp spike in the blue line.
In contrast, CARE strikes a balance:
CARE-0.3B-8B achieves an average score of 65.21, trailing the fully symmetric 8B giant (65.63) by only 0.6\%, yet requires 27$\times$ fewer online inference parameters (0.3B vs 8.2B).
This confirms that CARE effectively scales performance via the offline document encoder without increasing online latency.

\begin{table}[t]
\centering
\small
\setlength{\tabcolsep}{3.5pt} 
\resizebox{\columnwidth}{!}{%
\begin{tabular}{lccccc}
\toprule
\multirow{2}{*}{\textbf{Model}} & \multirow{2}{*}{\textbf{Asym}} & \multirow{2}{*}{\textbf{Params}} & \textbf{Retrieval} & \textbf{Rerank} & \multirow{2}{*}{\textbf{Avg}} \\
 & & & nDCG@10 & MAP@10 & \\
\midrule
KALE            & $\checkmark$  &0.3B / 4B& 42.67 & 67.42 & 55.05 \\
~\citet{wang2023query}     & $\checkmark$  &0.3B / 4B& 39.99 & 66.26 & 53.13 \\
ScalingNote     & $\checkmark$  &0.3B / 4B& 34.81 & 64.17 & 49.49 \\
Distill-from-4B & $\times$      &0.3B & \underline{54.68} & \underline{70.76} & \underline{62.72} \\
\textbf{CARE-0.3B-4B} & $\checkmark$  &0.3B / 4B& \textbf{55.91} & \textbf{72.84} & \textbf{64.38} \\
\bottomrule
\end{tabular}
}
\caption{\textbf{Ablation study on alternative approaches to efficient retrieval.} Params indicate parameter counts for Query / Document encoders.}
\label{tab:efficient_retr}

\end{table}

\subsection{Comparison with Efficient Retrieval Baselines (RQ3)}

We further compare our two-stage asymmetric training framework against several alternative symmetric or asymmetric approaches to efficient retrieval (Table~\ref{tab:efficient_retr}, implementation details are in ~\ref{app:efficient_baselines}).
CARE consistently outperforms all baselines. Asymmetric approaches like KALE and \citet{wang2023query} underperform (55.05 and 53.13), likely because their distillation designs originally tailored for encoder-only models transfer poorly to decoder-based architectures. 

Moreover, CARE surpasses the distillation baseline~\citep{ren2021rocketqav2} (distillation 62.72 vs. ours 64.38). While distillation attempts to transfer capabilities via soft labels (KL-divergence), it implies an inherent information loss. In contrast, our framework allows the query encoder to directly interact with teacher document encoder during inference, effectively bridging the information gap.

\subsection{Ablation Study (RQ4)}

We conduct ablation study on CARE-0.3B-4B except specially mentioned. To better reflect downstream applications, we report performance on both Retrieval and Reranking tasks.

\subsubsection{Training Design}

\begin{table}[t]
\centering
\small
\setlength{\tabcolsep}{4pt}
\resizebox{0.9\columnwidth}{!}{
\begin{tabular}{lccc}
\toprule
\multirow{2}{*}{\textbf{Setting}} & \textbf{Retrieval} & \textbf{Rerank} & \multirow{2}{*}{\textbf{Avg}} \\
 & nDCG@10 & MAP@10 & \\
\midrule
\multicolumn{4}{c}{\textit{Independent Initialization}} \\
\midrule
 w/o query init& 50.46& 68.85&59.66\\
 w/o doc init& 37.30& 63.21&50.26  \\
\midrule
\multicolumn{4}{c}{\textit{Asymmetric Stage}} \\
\midrule
w/o query align& 35.34& 66.79& 51.07  \\
w/o joint fine-tuning& 42.69& 68.28& 55.49\\
\midrule
\multicolumn{4}{c}{\textit{Loss Design (Asymmetric Stage \uppercase\expandafter{\romannumeral1})}} \\
\midrule
w/o MSE            & 55.19& 71.94& 63.57 \\
w/o Contrastive    & \underline{55.48}& \underline{72.58}& \underline{64.03} \\
\midrule
\textbf{Full Model} & \textbf{55.91}& \textbf{72.84}& \textbf{64.38} \\
\bottomrule
\end{tabular}
}
\caption{Ablation study on training stages and loss functions.}
\label{tab:ablation_stages}
\end{table}

We conduct experiments on the contribution of different components in our asymmetric training framework. Results are in Table~\ref{tab:ablation_stages}.

For Independent Initialization, removing either query or document encoder initialization  leads to severe performance degradation (\textit{w/o query init} scores: 59.66 and \textit{w/o doc init} scores: 50.26 vs. 64.38 for full model). This underscores the necessity of independent pre-training, where robust symmetric backbones serve as superior initializations for asymmetric alignment.

For Asymmetric Stage, skipping the query alignment stage (\textit{w/o query align} scores: 51.07) or the joint fine-tuning stage (\textit{w/o joint fine-tuning} scores: 55.49) results in clear performance drops (full model 64.38). This validates that the query alignment stage effectively bridges the semantic gap between student and teacher, while joint finetuning optimizes both encoders to downstream retrieval. Note that our \textit{w/o query align} configuration is close to HotelMatch~\citep{askari2025hotelmatch}, though not identical:
HotelMatch applies a linear projection to up-project the small-LM query embeddings to the document encoder dimension and uses separate learning rates for the two encoders, whereas we remove the projection layer and use a single learning rate since we use LoRA to fine-tune our document encoder.

Finally, we study the loss design in the query align stage. Removing either MSE or Asym-InfoNCE contrastive loss weakens performance. The full model, combining both, consistently achieves the best results. This indicates that both objectives are complementary for effective embedding space alignment.

\begin{table}[t]
\centering
\setlength{\tabcolsep}{3.5pt} 
\resizebox{\columnwidth}{!}{%
\begin{tabular}{l l ccc} 
\toprule
\multirow{2}{*}{\textbf{Training Phase}} & \multirow{2}{*}{\textbf{Stage-\uppercase\expandafter{\romannumeral1} Strategy}} & \textbf{Retrieval} & \textbf{Rerank} & \multirow{2}{*}{\textbf{Avg}} \\ &
 & nDCG@10 & MAP@10 & \\
\midrule
\multirow{2}{*}{Stage-\uppercase\expandafter{\romannumeral1} Only} 
    & Supervised (Labeled)       & 44.83 & 69.69 & 57.26 \\
    & Self-Contrastive (Ours)    & 42.69 & 68.28 & 55.49 \\
\midrule
\multirow{2}{*}{\shortstack[l]{Stage-\uppercase\expandafter{\romannumeral1} $\rightarrow$ Stage-\uppercase\expandafter{\romannumeral2}\\(Full Pipeline)}} 
    & Supervised (Labeled)       & \underline{49.95} & \underline{71.02} & \underline{60.49} \\
    & Self-Contrastive (Ours)    & \textbf{55.91} & \textbf{72.84} & \textbf{64.38} \\
\bottomrule
\end{tabular}%
}
\caption{
    \textbf{Impact of Stage-\uppercase\expandafter{\romannumeral1} Strategy.} 
    We compare using Supervised data vs. our Self-Contrastive approach during Stage-\uppercase\expandafter{\romannumeral1} alignment, followed by identical Stage-\uppercase\expandafter{\romannumeral2} fine-tuning.
}
\label{tab:stage1_data_ablation}
\end{table}

\subsubsection{Impact of Self-Contrastive Alignment}
We investigate the necessity of the unsupervised Self-Contrastive alignment task in Stage-\uppercase\expandafter{\romannumeral1} (Section~\ref{subsec:stage1}) by comparing it against using supervised contrastive fine-tuning data.
As detailed in Table~\ref{tab:stage1_data_ablation}, using supervised data in Stage-\uppercase\expandafter{\romannumeral1} yields higher immediate metrics compared to our self-contrastive approach (57.26 vs. 55.49). 
However, the trend reverses significantly after the Stage-\uppercase\expandafter{\romannumeral2} end-to-end fine-tuning: models initialized via our self-contrastive alignment achieve a decisive performance gain (64.38), surpassing those pre-trained with labeled data (60.49).

Early supervision in Stage-\uppercase\expandafter{\romannumeral1} likely causes \textit{premature convergence} to local minima. Conversely, the self-contrastive strategy provides a robust unsupervised alignment, establishing a superior foundation for subsequent fine-tuning.

\section{Conclusion}

In this work, we introduce CMedTEB, a new benchmark for Chinese medical text embedding, and propose \asyembedder, an asymmetric model designed for efficient, low-latency medical retrieval. Our architecture, which pairs a lightweight query encoder with a powerful document encoder via a two-stage training strategy, achieves state-of-the-art performance on CMedTEB. By releasing the benchmark, models, and training pipeline, we provide both a practical solution for real-world medical retrieval systems and a foundation for future research in domain-specific embedding learning. Our future work will include exploring more effective strategies for asymmetric alignment.

\section*{Limitations}

Our work presents a novel asymmetric alignment framework, yet it comes with certain limitations. First, regarding the architecture, the performance of our query encoder is inherently upper-bounded by the representational quality of the document encoder. Second, our work is primarily concentrated on the Chinese medical domain. While this setting presents significant challenges due to specialized terminology, the generalizability of our strategy to other languages or broader open-domain retrieval tasks remains to be verified. Finally, concerning the CMedTEB benchmark, although it utilizes a sophisticated multi-LLM pipeline for data construction, the annotation accuracy remains dependent on the capabilities of LLMs. However, this limitation may be mitigated by the advancement of LLMs.

\section*{Ethical considerations}

This research has been approved by the National Technology Ethics (Review) Committee, which is an IRB-equivalent ethics committee. We strictly adhered to ethical guidelines regarding data collection and privacy. User queries were sourced from participants who explicitly consented to a user experience improvement program for non-commercial research. Medical documents were crawled from publicly accessible, non-paywalled websites (e.g., XunYiWenYao\footnote{\url{https://www.xywy.com/}}) in compliance with robots.txt protocols and applicable copyright regulations. Both queries and documents were strictly anonymized, and details on the anonymization process are provided in Appendix \ref{app:medteb_anonymize}. We emphasize that these resources are for research purposes only and require rigorous validation before clinical deployment. Expert re-annotation was performed by clinicians from a partner tertiary hospital, compensated at institution-approved rates via official project budgets. CMedTEB is released under a CC BY-NC-SA 4.0 license, with model cards explicitly disclaiming diagnostic utility to ensure strict non-commercial, research-only usage.

\bibliography{custom}

@article{chen2024bge,
  title={Bge m3-embedding: Multi-lingual, multi-functionality, multi-granularity text embeddings through self-knowledge distillation},
  author={Chen, Jianlv and Xiao, Shitao and Zhang, Peitian and Luo, Kun and Lian, Defu and Liu, Zheng},
  journal={arXiv preprint arXiv:2402.03216},
  year={2024}
}

@article{lee2024nv,
  title={Nv-embed: Improved techniques for training llms as generalist embedding models},
  author={Lee, Chankyu and Roy, Rajarshi and Xu, Mengyao and Raiman, Jonathan and Shoeybi, Mohammad and Catanzaro, Bryan and Ping, Wei},
  journal={arXiv preprint arXiv:2405.17428},
  year={2024}
}

@article{li2024making,
  title={Making text embedders few-shot learners},
  author={Li, Chaofan and Qin, MingHao and Xiao, Shitao and Chen, Jianlyu and Luo, Kun and Shao, Yingxia and Lian, Defu and Liu, Zheng},
  journal={arXiv preprint arXiv:2409.15700},
  year={2024}
}

@article{li2023towards,
  title={Towards general text embeddings with multi-stage contrastive learning},
  author={Li, Zehan and Zhang, Xin and Zhang, Yanzhao and Long, Dingkun and Xie, Pengjun and Zhang, Meishan},
  journal={arXiv preprint arXiv:2308.03281},
  year={2023}
}

@article{zhang2025qwen3,
  title={Qwen3 Embedding: Advancing Text Embedding and Reranking Through Foundation Models},
  author={Zhang, Yanzhao and Li, Mingxin and Long, Dingkun and Zhang, Xin and Lin, Huan and Yang, Baosong and Xie, Pengjun and Yang, An and Liu, Dayiheng and Lin, Junyang and others},
  journal={arXiv preprint arXiv:2506.05176},
  year={2025}
}

@article{izacard2021unsupervised,
  title={Unsupervised dense information retrieval with contrastive learning},
  author={Izacard, Gautier and Caron, Mathilde and Hosseini, Lucas and Riedel, Sebastian and Bojanowski, Piotr and Joulin, Armand and Grave, Edouard},
  journal={arXiv preprint arXiv:2112.09118},
  year={2021}
}

@inproceedings{xiao2024c,
  title={C-pack: Packed resources for general chinese embeddings},
  author={Xiao, Shitao and Liu, Zheng and Zhang, Peitian and Muennighoff, Niklas and Lian, Defu and Nie, Jian-Yun},
  booktitle={Proceedings of the 47th international ACM SIGIR conference on research and development in information retrieval},
  pages={641--649},
  year={2024}
}

@article{wang2022text,
  title={Text embeddings by weakly-supervised contrastive pre-training},
  author={Wang, Liang and Yang, Nan and Huang, Xiaolong and Jiao, Binxing and Yang, Linjun and Jiang, Daxin and Majumder, Rangan and Wei, Furu},
  journal={arXiv preprint arXiv:2212.03533},
  year={2022}
}

@article{wang2023query,
  title={Query encoder distillation via embedding alignment is a strong baseline method to boost dense retriever online efficiency},
  author={Wang, Yuxuan and Lyu, Hong},
  journal={arXiv preprint arXiv:2306.11550},
  year={2023}
}

@article{campos2023quick,
  title={Quick dense retrievers consume kale: Post training kullback leibler alignment of embeddings for asymmetrical dual encoders},
  author={Campos, Daniel and Magnani, Alessandro and Zhai, ChengXiang},
  journal={arXiv preprint arXiv:2304.01016},
  year={2023}
}

@article{huang2024scalingnote,
  title={Scalingnote: Scaling up retrievers with large language models for real-world dense retrieval},
  author={Huang, Suyuan and Zhang, Chao and Wu, Yuanyuan and Zhang, Haoxin and Wang, Yuan and Wang, Maolin and Cao, Shaosheng and Xu, Tong and Zhao, Xiangyu and Qin, Zengchang and others},
  journal={arXiv preprint arXiv:2411.15766},
  year={2024}
}

@article{askari2025hotelmatch,
  title={HotelMatch-LLM: Joint Multi-Task Training of Small and Large Language Models for Efficient Multimodal Hotel Retrieval},
  author={Askari, Arian and Stergiadis, Emmanouil and Gusev, Ilya and Beladev, Moran},
  journal={arXiv preprint arXiv:2506.07296},
  year={2025}
}

@article{muennighoff2022mteb,
  title={Mteb: Massive text embedding benchmark},
  author={Muennighoff, Niklas and Tazi, Nouamane and Magne, Lo{\"\i}c and Reimers, Nils},
  journal={arXiv preprint arXiv:2210.07316},
  year={2022}
}

@article{zhang2017chinese,
  title={Chinese Medical Question Answer Matching Using End-to-End Character-Level Multi-Scale CNNs},
  author={Zhang, Sheng and Zhang, Xin and Wang, Hui and Cheng, Jiajun and Li, Pei and Ding, Zhaoyun},
  journal={Applied Sciences},
  volume={7},
  number={8},
  pages={767},
  year={2017},
  publisher={Multidisciplinary Digital Publishing Institute}
}

@ARTICLE{8548603, 
author={S. Zhang and X. Zhang and H. Wang and L. Guo and S. Liu}, 
journal={IEEE Access}, 
title={Multi-Scale Attentive Interaction Networks for Chinese Medical Question Answer Selection}, 
year={2018}, 
volume={6}, 
number={}, 
pages={74061-74071}, 
doi={10.1109/ACCESS.2018.2883637}, 
ISSN={2169-3536}, 
month={},}

@inproceedings{Long2022MultiCPRAM,
  author    = {Dingkun Long and Qiong Gao and Kuan Zou and Guangwei Xu and Pengjun Xie and Ruijie Guo and Jian Xu and Guanjun Jiang and Luxi Xing and Ping Yang},
  title     = {Multi-CPR: {A} Multi Domain Chinese Dataset for Passage Retrieval},
  booktitle = {{SIGIR}},
  pages     = {3046--3056},
  publisher = {{ACM}},
  year      = {2022}
}

@article{reimers2019sentence,
  title={Sentence-bert: Sentence embeddings using siamese bert-networks},
  author={Reimers, Nils and Gurevych, Iryna},
  journal={arXiv preprint arXiv:1908.10084},
  year={2019}
}

@article{lewis2020retrieval,
  title={Retrieval-augmented generation for knowledge-intensive nlp tasks},
  author={Lewis, Patrick and Perez, Ethan and Piktus, Aleksandra and Petroni, Fabio and Karpukhin, Vladimir and Goyal, Naman and K{\"u}ttler, Heinrich and Lewis, Mike and Yih, Wen-tau and Rockt{\"a}schel, Tim and others},
  journal={Advances in neural information processing systems},
  volume={33},
  pages={9459--9474},
  year={2020}
}

@article{liu2024deepseek,
  title={Deepseek-v3 technical report},
  author={Liu, Aixin and Feng, Bei and Xue, Bing and Wang, Bingxuan and Wu, Bochao and Lu, Chengda and Zhao, Chenggang and Deng, Chengqi and Zhang, Chenyu and Ruan, Chong and others},
  journal={arXiv preprint arXiv:2412.19437},
  year={2024}
}

@article{hurst2024gpt,
  title={Gpt-4o system card},
  author={Hurst, Aaron and Lerer, Adam and Goucher, Adam P and Perelman, Adam and Ramesh, Aditya and Clark, Aidan and Ostrow, AJ and Welihinda, Akila and Hayes, Alan and Radford, Alec and others},
  journal={arXiv preprint arXiv:2410.21276},
  year={2024}
}

@article{guo2025seed1,
  title={Seed1. 5-vl technical report},
  author={Guo, Dong and Wu, Faming and Zhu, Feida and Leng, Fuxing and Shi, Guang and Chen, Haobin and Fan, Haoqi and Wang, Jian and Jiang, Jianyu and Wang, Jiawei and others},
  journal={arXiv preprint arXiv:2505.07062},
  year={2025}
}

@article{xiao2022retromae,
  title={RetroMAE: Pre-training retrieval-oriented language models via masked auto-encoder},
  author={Xiao, Shitao and Liu, Zheng and Shao, Yingxia and Cao, Zhao},
  journal={arXiv preprint arXiv:2205.12035},
  year={2022}
}

@article{oord2018representation,
  title={Representation learning with contrastive predictive coding},
  author={Oord, Aaron van den and Li, Yazhe and Vinyals, Oriol},
  journal={arXiv preprint arXiv:1807.03748},
  year={2018}
}

@article{kusupati2022matryoshka,
  title={Matryoshka representation learning},
  author={Kusupati, Aditya and Bhatt, Gantavya and Rege, Aniket and Wallingford, Matthew and Sinha, Aditya and Ramanujan, Vivek and Howard-Snyder, William and Chen, Kaifeng and Kakade, Sham and Jain, Prateek and others},
  journal={Advances in Neural Information Processing Systems},
  volume={35},
  pages={30233--30249},
  year={2022}
}

@article{li2024conan,
  title={Conan-embedding: General text embedding with more and better negative samples},
  author={Li, Shiyu and Tang, Yang and Chen, Shizhe and Chen, Xi},
  journal={arXiv preprint arXiv:2408.15710},
  year={2024}
}

@article{zhang2024jasper,
  title={Jasper and stella: distillation of sota embedding models},
  author={Zhang, Dun and Li, Jiacheng and Zeng, Ziyang and Wang, Fulong},
  journal={arXiv preprint arXiv:2412.19048},
  year={2024}
}

@article{ren2021rocketqav2,
  title={RocketQAv2: A joint training method for dense passage retrieval and passage re-ranking},
  author={Ren, Ruiyang and Qu, Yingqi and Liu, Jing and Zhao, Wayne Xin and She, Qiaoqiao and Wu, Hua and Wang, Haifeng and Wen, Ji-Rong},
  journal={arXiv preprint arXiv:2110.07367},
  year={2021}
}

@article{chen2024air,
  title={Air-bench: Automated heterogeneous information retrieval benchmark},
  author={Chen, Jianlyu and Wang, Nan and Li, Chaofan and Wang, Bo and Xiao, Shitao and Xiao, Han and Liao, Hao and Lian, Defu and Liu, Zheng},
  journal={arXiv preprint arXiv:2412.13102},
  year={2024}
}

@inproceedings{thomas2024large,
  title={Large language models can accurately predict searcher preferences},
  author={Thomas, Paul and Spielman, Seth and Craswell, Nick and Mitra, Bhaskar},
  booktitle={Proceedings of the 47th International ACM SIGIR Conference on Research and Development in Information Retrieval},
  pages={1930--1940},
  year={2024}
}

@inproceedings{xu-etal-2024-bmretriever,
    title = "{BMR}etriever: Tuning Large Language Models as Better Biomedical Text Retrievers",
    author = "Xu, Ran  and
      Shi, Wenqi  and
      Yu, Yue  and
      Zhuang, Yuchen  and
      Zhu, Yanqiao  and
      Wang, May Dongmei  and
      Ho, Joyce C.  and
      Zhang, Chao  and
      Yang, Carl",
    editor = "Al-Onaizan, Yaser  and
      Bansal, Mohit  and
      Chen, Yun-Nung",
    booktitle = "Proceedings of the 2024 Conference on Empirical Methods in Natural Language Processing",
    month = nov,
    year = "2024",
    address = "Miami, Florida, USA",
    publisher = "Association for Computational Linguistics",
    url = "https://aclanthology.org/2024.emnlp-main.1241/",
    doi = "10.18653/v1/2024.emnlp-main.1241",
    pages = "22234--22254"
}

@article{10.1093/bioinformatics/btad651,
    author = {Jin, Qiao and Kim, Won and Chen, Qingyu and Comeau, Donald C and Yeganova, Lana and Wilbur, W John and Lu, Zhiyong},
    title = {MedCPT: Contrastive Pre-trained Transformers with large-scale PubMed search logs for zero-shot biomedical information retrieval},
    journal = {Bioinformatics},
    volume = {39},
    number = {11},
    pages = {btad651},
    year = {2023},
    month = {11},
    issn = {1367-4811},
    doi = {10.1093/bioinformatics/btad651},
    url = {https://doi.org/10.1093/bioinformatics/btad651},
    eprint = {https://academic.oup.com/bioinformatics/article-pdf/39/11/btad651/52799559/btad651.pdf},
}

@article{fleiss1971measuring,
  title={Measuring nominal scale agreement among many raters.},
  author={Fleiss, Joseph L},
  journal={Psychological bulletin},
  volume={76},
  number={5},
  pages={378},
  year={1971},
  publisher={American Psychological Association}
}

@inproceedings{li-etal-2025-automir,
    title = "{A}uto{MIR}: Effective Zero-Shot Medical Information Retrieval without Relevance Labels",
    author = "Li, Lei  and
      Zhang, Xiangxu  and
      Zhou, Xiao  and
      Liu, Zheng",
    editor = "Christodoulopoulos, Christos  and
      Chakraborty, Tanmoy  and
      Rose, Carolyn  and
      Peng, Violet",
    booktitle = "Findings of the Association for Computational Linguistics: EMNLP 2025",
    month = nov,
    year = "2025",
    address = "Suzhou, China",
    publisher = "Association for Computational Linguistics",
    url = "https://aclanthology.org/2025.findings-emnlp.1305/",
    doi = "10.18653/v1/2025.findings-emnlp.1305",
    pages = "24028--24047",
    ISBN = "979-8-89176-335-7"
}

@inproceedings{fan2025medeureka,
  title={MedEureka: A Medical Domain Benchmark for Multi-Granularity and Multi-Data-Type Embedding-Based Retrieval},
  author={Fan, Yongqi and Wang, Nan and Xue, Kui and Liu, Jingping and Ruan, Tong},
  booktitle={Findings of the Association for Computational Linguistics: NAACL 2025},
  pages={2825--2851},
  year={2025}
}

@article{zhang2026stable,
  title={Stable-RAG: Mitigating Retrieval-Permutation-Induced Hallucinations in Retrieval-Augmented Generation},
  author={Zhang, Qianchi and Zhang, Hainan and Pang, Liang and Zheng, Hongwei and Zheng, Zhiming},
  journal={arXiv preprint arXiv:2601.02993},
  year={2026}
}

@article{lin2025causal2vec,
  title={Causal2Vec: Improving Decoder-only LLMs as Versatile Embedding Models},
  author={Lin, Ailiang and Li, Zhuoyun and Funakoshi, Kotaro and Okumura, Manabu},
  journal={arXiv preprint arXiv:2507.23386},
  year={2025}
}

\newpage
\appendix

\appendix
\section{Details of CMedTEB Benchmark}
\label{app:medteb_detail}

\subsection{Instruction}
Table~\ref{tab:task_instruction_templates} presents the instruction used on CMedTEB benchmarks.
\begin{table*}[t] 
    \centering
    \small 
    
    \resizebox{\textwidth}{!}{ 
        \begin{tabular}{l p{12cm}} 
        \toprule
        \textbf{Task Name} & \textbf{Instruction Template} \\
        \midrule
        CMedQAv1-reranking & Based on a Chinese medical question, evaluate and rank the medical information that provide answers to the question. \\
        CMedQAv2-reranking & Based on a Chinese medical question, evaluate and rank the medical information that provide answers to the question. \\
        CMedTEB-Retrieval & Given a Chinese medical question, retrieve medical documents that answer the question. \\
        CMedTEB-Rerank & Based on a Chinese medical question, evaluate and rank the medical information that provide answers to the question. \\
        CMedTEB-STS & Retrieve semantically similar text. \\
        \bottomrule
        \end{tabular}
    }
    \caption{Instruction used on CMedTEB benchmarks}
    \label{tab:task_instruction_templates}
\end{table*}

\subsection{Statistics}
\label{app:medteb_stats_app}
{For detailed statistics of the CMedTEB datasets, please refer to Tables~\ref{tab:medteb_full_stats}. 
To measure sequence lengths, we utilize the \texttt{tiktoken} tokenizer with the \texttt{cl100k\_base} encoding scheme to count tokens for queries and corpus documents.}

\begin{table*}[t]
\centering
\resizebox{\textwidth}{!}{
\begin{tabular}{ll c rr rr r}
\toprule
\textbf{Split} & \textbf{Dataset} & \textbf{Type} & \textbf{\# Queries} & \textbf{Avg. $|Q|$} & \textbf{\# Corpus / Cand.} & \textbf{Avg. $|D|$} & \textbf{Avg. Pos / Pos. Ratio} \\
\midrule
\multirow{3}{*}{\textbf{Train}} 
 & Retrieval / Rerank & Mixed   & 20,000 & 21.24 & 229,457 & 470.90 & - \\
 & CMedQA             & Rerank  & 50,000 & 66.76 & 196,902 & 134.22 & - \\
 & STS                & STS     & 10,000 & 23.52 & 24,906  & 29.95  & - \\
\midrule
\multirow{5}{*}{\textbf{Test}} 
 & CMedTEB-Retrieval   & Retr.   & 734    & 20.68 & 229,457 & 470.90 & 8.43 \\
 & CMedTEB-Rerank      & Rerank  & 1,128  & 18.52 & 27.83   & 502.75 & 7.83 \\
 & CMedQA-v1          & Rerank  & 1,000  & 75.58 & 100.00  & 143.03 & 1.93 \\
 & CMedQA-v2          & Rerank  & 1,000  & 66.98 & 100.00  & 135.99 & 1.91 \\
 & CMedTEB-STS         & STS     & 5,000  & 35.45 & -       & -      & 47.92\% \\
\bottomrule
\end{tabular}
}
\caption{
\textbf{Detailed Statistics of CMedTEB Datasets.} 
We aggregate statistics for both Training and Test splits across Retrieval, Reranking, and STS tasks. 
\textit{Avg. $|Q|$} and \textit{Avg. $|D|$} denote the average lengths of queries and documents, respectively. 
For STS, the last column indicates the ratio of positive pairs.
}
\label{tab:medteb_full_stats}
\end{table*}

\subsection{Anonymization steps of CMedTEB.}
\label{app:medteb_anonymize}
All user queries and web documents were processed as follows. 1) Automated PII (personally identifiable information) Detection: We deployed an offline, locally hosted large language model to detect and mask potential PII, including names, locations, phone numbers, and ID numbers. 2) Rule-based Validation: After initial masking, we applied a rule-based validation module to scan residual digits, and keywords. 3) Human Checks: 1\% of anonymized data were checked by human, and no re-identifiable content found.

\subsection{CMedTEB Case Examples}
\label{app:medteb_examples}

We present representative cases from CMedTEB tasks in Table~\ref{tab:medteb-retri-example}, Table~\ref{tab:medteb-rerank-example} and Table~\ref{tab:medteb-sts-example}.
\begin{table*}[t]
    \centering
    \begin{small}
    \begin{tabular}{p{0.95\linewidth} }
        \toprule
        \textbf{Query} \\
        \begin{CJK}{UTF8}{gkai} 肾结石如何判断是酸性还是碱性结石？
        \end{CJK} \\
        \textit{How to ascertain whether a renal calculus is acidic or alkaline in composition?}  \\
        \midrule
        \textbf{Positive example} \\
        \begin{CJK}{UTF8}{gkai}咋知道肾结石是有酸性碱性引起 
        
        病情分析：一般通过尿检判断肾结石是酸性的还是碱性的，可以到本地正规医院做尿液，酸碱度检查也可以观察一下pH值的变化，然后再明确一下尿液的酸碱度。如果怀疑身体有肾结石的症状，可以到正规医院做影像学检查检查一下大小。如果结石比较大的话，一定要及时到医院做激光碎石治疗。
        \end{CJK} \\
        \textit{How can I tell whether a kidney stone is related to acidic or alkaline urine? Clinical assessment: In general, urinalysis is used to determine whether a renal calculus is associated with acidic or alkaline urine. You can have a urine pH test at a qualified local hospital and monitor the pH value to establish urinary acidity or alkalinity. If kidney stone symptoms are suspected, undergo imaging studies to assess the stone size. If the calculus is relatively large, timely laser lithotripsy is recommended.} \\
        \bottomrule
    \end{tabular}
    \end{small}
    \caption{
    CMedTEB-Retrieval example. 
    }
    \label{tab:medteb-retri-example}
\end{table*}

\begin{table*}[t]
    \centering
    \begin{small}
    \begin{tabular}{p{0.95\linewidth} }
        \toprule
        \textbf{Query} \\
        \begin{CJK}{UTF8}{gkai} 耳鸣需要吃什么药？
        \end{CJK} \\
        \textit{Which medications are indicated for tinnitus?}  \\
        \midrule
        \textbf{Positive example} \\
        \begin{CJK}{UTF8}{gkai}耳鸣的药有哪些
        
        病情分析：耳鸣常用的药物有，1.盐酸氟桂利嗪胶囊、尼莫地平等，用于改善耳蜗的供血，扩张耳蜗血管。2. 三磷酸干、辅酶A、甲钴胺等，用于改善耳道的代谢功能，可以促进耳部的新陈代谢，清理耳道杂质。3.卡马西平、路硝西泮等，用于抗惊厥，能够缓解耳朵受到刺激造成的耳鸣。4. 抗生素、红霉素、万古霉素等，这些药物含有非类固醇消炎药物，可以给耳道涂抹起到消炎的作用，以此来缓解耳鸣。
        \end{CJK} \\
        \textit{What medications are available for tinnitus? Clinical assessment: Commonly used drugs include flunarizine hydrochloride capsules and nimodipine to improve cochlear perfusion by dilating cochlear vessels; adenosine triphosphate (ATP), coenzyme A, and mecobalamin (methylcobalamin) to enhance metabolic function of the auditory pathway, promote aural metabolism, and help clear debris from the ear canal; carbamazepine and clonazepam as anticonvulsants to relieve tinnitus triggered by neural irritation; and antibiotics such as erythromycin and vancomycin, as well as nonsteroidal anti‑inflammatory agents, which can be applied to the ear canal for anti‑inflammatory effects to help alleviate tinnitus.} \\
        \midrule
        \textbf{Negative example} \\
        \begin{CJK}{UTF8}{gkai}吃补肾的药怎么耳鸣呢
        
        病情分析：患者是由于肾阴亏虚而引起的上火症状，进而导致患者出现耳鸣。首先，患者应该服用一些滋阴补肾的药物来进行补肾，比如六味地黄丸或者知柏地黄丸。等到患者的肾虚得到一定的恢复之后，耳鸣的症状也会逐渐的消失。另外，患者可以搭配服用一些清热泻火的药物来进行治疗。
        \end{CJK} \\
        \textit{Why would taking kidney‑tonifying medicine lead to tinnitus? Clinical assessment: From a traditional Chinese medicine perspective, the patient’s tinnitus is due to kidney‑yin deficiency with endogenous heat, which precipitates tinnitus. It is advisable to use yin‑nourishing, kidney‑tonifying formulas such as Liuwei Dihuang Wan or Zhibai Dihuang Wan. As the kidney deficiency improves, the tinnitus should gradually resolve. In addition, heat‑clearing and fire‑purging agents can be used concomitantly.} \\
        \bottomrule
    \end{tabular}
    \end{small}
    \caption{
    CMedTEB-Rerank example. 
    }
    \label{tab:medteb-rerank-example}
\end{table*}

\begin{table*}[t]
    \centering
    \scalebox{1.00}{
    \begin{small}
    \begin{tabular}{p{0.95\linewidth} }
        \toprule
        \textbf{Sentence1} \\
        \begin{CJK}{UTF8}{gkai} 碳酸氢钠片是否会引起头皮痒
        \end{CJK} \\
        \textit{Do sodium bicarbonate tablets cause scalp itching?}  \\
        \midrule
        \textbf{Sentence2} \\
        \begin{CJK}{UTF8}{gkai}服用小苏打片是否可能导致头皮发痒？
        \end{CJK} \\
        \textit{Could taking baking soda tablets lead to an itchy scalp?} \\
        \bottomrule
    \end{tabular}
    \end{small}
    }
    \caption{
    CMedTEB-STS example. 
    }
    \label{tab:medteb-sts-example}
\end{table*}

\subsection{Zero-shot results on CMedTEB.}
Table~\ref{tab:zero_shot_medteb} presents the full zero-shot performance of all evaluated models across individual CMedTEB tasks. Results show significant performance gaps between general-domain embedders and the medical-specific retrieval challenge. Note that most of baselines have already trained on CMedQA train dataset before.

\label{app:zero-shot-medteb}
\begin{table*}[th]
\small
\centering
\setlength{\tabcolsep}{3pt}
\resizebox{\textwidth}{!}{
\begin{tabular}{l|c|cc|c|ccc|c}
\toprule
\multirow{2}{*}{\textbf{Model}} & \multirow{2}{*}{\textbf{Param.}} & \textbf{CMedv1} & \textbf{CMedv2} & \multirow{2}{*}{\textbf{Avg CMed}} & \textbf{Retr.} & \textbf{Rerank} & \textbf{STS} & \multirow{2}{*}{\textbf{Avg. New}} \\
 & & MAP@10 & MAP@10 & & nDCG@10 & MAP@10 & Pearson & \\
\midrule
gte-multilingual-base         &305M& 86.11 & 87.40 & 86.76 & 47.80 & 61.51 & 72.39 & 60.57 \\
gte-base-zh                   &102M& 86.79 & 87.20 & 86.99 & 44.18 & 58.40 & 75.07 & 59.22 \\
gte-large-zh                  &326M& 86.09 & 86.46 & 86.28 & 29.75 & 53.70 & 68.02 & 50.49 \\
gte-Qwen2-1.5B-instruct                &1.78B& 88.16 & 88.12 & 88.14 & 45.14 & 58.99 & 76.81 & 60.31 \\
gte-Qwen2-7B-instruct                  &7.61B& 88.20 & \underline{89.31}& \underline{88.76}& 40.94 & 61.07 & 72.67 & 58.23 \\
bge-small-zh-v1.5             &24M& 77.40 & 79.86 & 78.63 & 35.22 & 55.39 & 57.87 & 49.49 \\
bge-base-zh-v1.5              &102M& 80.47 & 84.88 & 82.68 & 33.11 & 53.56 & 67.45 & 51.37 \\
bge-large-zh-v1.5             &326M& 83.45 & 85.44 & 84.45 & 43.05 & 58.31 & 71.90 & 57.75 \\
bge-m3                        &568M& 77.71 & 79.19 & 78.45 & 41.14 & 57.68 & 63.67 & 54.16 \\
Conan-embedding-v1                      &326M& \textbf{91.39} & \textbf{89.72} & \textbf{90.56} & 41.60 & 61.89 & 72.86 & 58.78 \\
stella-base-zh-v3-1792d       &102M& \underline{88.35}& 89.06 & 88.71 & 45.77 & 60.43 & 74.96 & 60.39 \\
Qwen3-Embedding-0.6B          &596M& 80.06 & 81.35 & 80.71 & 47.54 & 64.51 & 68.31 & 60.12 \\
Qwen3-Embedding-4B            &4.02B& 84.43 & 85.06 & 84.75 & \underline{50.14}& \textbf{66.67}& \textbf{76.49}& \underline{64.43}\\
Qwen3-Embedding-8B            &7.57B& 86.13 & 86.39 & 86.26 & \textbf{51.15} & \underline{66.31}& \underline{76.09}& \textbf{64.52} \\
\midrule
 Average performance && 84.62& 85.67& 85.15& 42.61& 59.89& 71.04&57.85\\
\midrule
  \multicolumn{8}{l}{Spearman Rank Correlation Coefficient (P-value)}&0.354 (0.215)\\
\bottomrule
\end{tabular}
}
\caption{Zero-shot results on CMedTEB (\%). Best results in \textbf{bold}.}
\label{tab:zero_shot_medteb}
\end{table*}

\subsection{Dataset Comparative Analysis}
\label{app:dataset-comparison}

We compare CMedTEB with CMIRB and MedEureka~\citep{fan2025medeureka} across four key dimensions. \textbf{False Negative Handling:} CMIRB aggregates legacy datasets with annotation sparsity; MedEureka verifies original pairs but does not systematically retrieve other candidates. CMedTEB mitigates this by retrieving top candidates using multiple retrievers and re-annotating the entire pool. \textbf{Label Construction:} CMIRB uses ChatGPT for filtering; MedEureka employs GPT-4o with $\sim$10\% human correction. CMedTEB implements Multi-LLM Voting followed by Clinical Expert Verification, achieving Fleiss' Kappa of 0.731 and 93.3\% expert agreement. \textbf{Query Authenticity:} MedEureka relies partially on LLM-generated queries, while CMedTEB derives from real-world user logs preserving authentic distributions. \textbf{Task Scope:} CMIRB and MedEureka focus on Retrieval only; CMedTEB extends to Reranking and STS.

\subsection{Validation of Consensus Annotation Pipeline}
\label{app:annotation_validation}

Unlike general domain retrieval, medical tasks demand high precision due to complex knowledge structures. To validate our strict consensus pipeline, we analyzed 100 random samples where three LLMs (GPT-4o, Kimi, Claude-3.5-Sonnet) disagreed. As shown in Table~\ref{tab:disagreement}, 50\% involve boundary relevance with multiple ambiguous intents, 38\% suffer from LLM hallucinations on complex terminology, and 12\% contain low-quality queries/documents. Including these would compromise reliability by increasing false negatives rather than contributing valid difficulty.

\begin{table*}[h]
\centering
\small
\begin{tabular}{@{}l p{6.5cm} c@{}}
\toprule
\textbf{Reason} & \textbf{Description} & \textbf{Ratio (\%)} \\
\midrule
Boundary relevance & Multiple or ambiguous intents make relevance labeling inherently difficult. & 50 \\
LLM hallucination & Limited medical knowledge causes misjudgment of complex terms or relationships. & 38 \\
Noise & Low-quality or incomplete queries/documents mislead into incorrect labels. & 12 \\
\bottomrule
\end{tabular}
\caption{Analysis of LLM disagreement reasons (N=100).}
\label{tab:disagreement}
\end{table*}

Consensus does not imply triviality. To verify CMedTEB retains inherent complexity, we sampled 100 queries and employed GPT-4o to annotate difficulty levels. As shown in Table~\ref{tab:difficulty}, only 26\% can be solved by lexical matching, while 53\% require semantic understanding and 21\% need deep inference involving medical knowledge or multi-hop logic. Furthermore, strong baselines exhibit a sharp performance drop from 85.15\% on CMedQA to 57.85\% on CMedTEB's new tasks (Section 3.2), confirming our pipeline filters ambiguity while preserving rigorous evaluation difficulty.

\begin{table*}[h]
\centering
\small
\begin{tabular}{@{}l p{6.5cm} c@{}}
\toprule
\textbf{Level} & \textbf{Description} & \textbf{Ratio (\%)} \\
\midrule
L1: Lexical & High keyword overlap between query and document. & 26 \\
L2: Semantic & Conceptual match without keyword overlap; requires dense retrieval. & 53 \\
L3: Hard & Deep inference requiring medical knowledge, cause-effect, or multi-hop logic. & 21 \\
\bottomrule
\end{tabular}
\caption{Difficulty distribution of CMedTEB samples (N=100).}
\label{tab:difficulty}
\end{table*}

\section{Training Data Construction Details}
\label{app:data_diversify}
\subsection{Data Construction}
\paragraph{Data Diversification.} We apply diversification for query and corpus independently. We first initialized a vector index seeded with 5{,}000 documents encoded by \texttt{gte-multilingual-base}. For each new candidate \(x\) (query or document), we retrieve top-\(k\) neighbors and discard \(x\) if more than \(n\) neighbors exceed similarity threshold \(t\); otherwise we insert \(x\). This is applied separately to queries and corpus, preserving diversity while removing near-duplicates.
We summarize the key parameters used during Data Diversification in Table~\ref{tab:data_construction_params}, where $k$ represents for top-$k$ retrieved relevant candidates from vector index, $t$ for similarity score threshold, and $n$ for maximum number of related documents. 
\begin{table}[h]
\centering
\small
\setlength{\tabcolsep}{10pt}
\begin{tabular}{lcc}
\toprule
\textbf{Parameter} & \textbf{Query} & \textbf{Document} \\
\midrule
$k$ (retrieved candidates) & 5 & 5 \\
$t$ (score threshold) & 0.85 & 0.78 \\
$n$ (maximum number) & 1 & 1 \\
\bottomrule
\end{tabular}
\caption{Key parameters used during data generation.}
\label{tab:data_construction_params}
\end{table}

\paragraph{LLM annotation.}
For each diversified query \(q\), we retrieve top-50 candidates from the diversified corpus and have GPT-4o assign a 5-point relevance score. From scored pools, we select positives and negatives to form triples, yielding 500K fine-tuning instances of triplets $\mathcal{T}=\{(q_i,\mathcal{P}_i,\mathcal{N}_i)\}$, where $\mathcal{P}_i$ is a list sampled from positives $P_i$ and $\mathcal{N}_i$ is a list sampled from negatives $N_i$.

\subsection{Ablation studies on Data Diversification}
To evaluate the effectiveness of our diversity-aware data curation pipeline, we conduct an ablation study on the role of query and document-side diversification on Medical-Embedder-base. All configurations use the same amount of training data.
As shown in Table~\ref{tab:diversify}, the full setting achieves the best performance, demonstrating that both query and document diversification are essential: the former ensures broad topic coverage, while the latter improves the reliability and difficulty of negative samples. This validates the importance of our diversity-aware curation strategy in building high-quality medical retrieval datasets.

\begin{table}[!ht]
\centering
\small
\begin{tabular}{lccc}
\toprule
\multirow{2}{*}{\textbf{Diversification Setting}} & \textbf{Retrieval} & \textbf{Rerank} & \multirow{2}{*}{\textbf{Avg}} \\
 & nDCG@10 & MAP@10 & \\
\midrule
w/o query, w/o doc   &51.17& 68.74&59.96\\
w/ query,  w/o doc   &\underline{52.23}& \underline{68.98}& \underline{60.61}\\
w/ query,  w/ doc    &\textbf{54.16}& \textbf{69.63}&\textbf{61.90}\\
\bottomrule
\end{tabular}
\caption{Impact of query and document diversification on retrieval performance.}

\label{tab:diversify}
\end{table}

\section{Implementation Details}
\subsection{Independent Initialization}
\label{app:independent_initialization}
\subsubsection{Query Encoder Training}
\label{app:query_training}
\paragraph{RetroMAE Pretrain.} We first adopt RetroMAE~\citep{xiao2022retromae} pretrain, which mask inputs differently in the encoder and a lightweight decoder; the encoder outputs sentence embeddings and the decoder reconstructs the original text via masked language modeling. This stage leverages a 60M unsupervised Medical Q\&A corpus.

\paragraph{Unsupervised Pretrain.} We perform contrastive unsupervised pretrain using InfoNCE loss~\citep{oord2018representation}:
\[
\mathcal{L}_{\text{InfoNCE}} = -\log \frac{
    \exp(\mathbf{q}^\top \mathbf{d}^+ / \tau)
}{
    \sum_{\mathbf{d} \in \mathcal{D}} \exp(\mathbf{q}^\top \mathbf{d} / \tau)
},
\]
where $\mathbf{q}, \mathbf{d}^+$ are embeddings of a matched (query, document) pair, $\mathcal{D}$ contains one positive and $|\mathcal{D}|-1$ negatives, and $\tau$ is a learnable temperature. We use the same unsupervised medical Q\&A corpus for RetroMAE pretraining, treating title–content pairs as positives and other documents within the same batch as in-batch negatives.

\paragraph{Supervised Finetuning.} The final stage fine-tunes the encoder on high quality fine-tuning datasets described in Section~\ref{subsec:data_construction} together with the training splits of \textbf{CMedTEB} (retrieval, reranking, CMedQA, and Synonym STS) using the InfoNCE loss. 

\subsubsection{Document Encoder Training}
\label{app:doc_training}
We fine-tune Qwen3-4B and Qwen3-8B and apply LoRA with rank=32, $\alpha=64$. We adopt Matryoshka Representation Learning (MRL)~\citep{kusupati2022matryoshka}, whose training objective aggregates the contrastive loss across this predefined set of dimensions. Specifically, the final loss is the average of the InfoNCE losses computed at each target dimension:
\begin{equation}
    \mathcal{L}_{\text{MRL}} = \frac{1}{|M|} \sum_{m \in M} \mathcal{L}_{\text{InfoNCE}}^{(m)},
\end{equation}
where \(M\) is the set of nested dimensions and \(\mathcal{L}_{\text{InfoNCE}}^{(m)}\) is the standard InfoNCE loss calculated using embeddings truncated to the first \(m\) dimensions.

\subsubsection{Impact of Pretraining on Query Encoder}
\label{app:query_pretrain}
We evaluate the impact of pretraining on the query encoder. 
As shown in Table~\ref{tab:ablation_pretraining}, combining RetroMAE and unsupervised domain pretraining achieves the best performance (54.16), outperforming ablated variants. 
This confirms that multi-stage pretraining enhances the encoder’s performance in medical retrieval. 

\begin{table}[ht]
\centering
\begin{tabular}{lr}
\toprule
\multirow{2}{*}{\textbf{Training Strategy}} & \textbf{Retrieval} \\
 & nDCG@10 \\
\midrule
Finetune only & 52.88 \\
RetroMAE + Finetune & 53.21 \\
RetroMAE + Unsup + Finetune & \textbf{54.16} \\
\bottomrule
\end{tabular}
\caption{Ablation study on pretraining strategies for Medical-Embedder-Base. 
Combining RetroMAE and unsupervised domain pretraining leads to the best retrieval performance.}
\label{tab:ablation_pretraining}
\end{table}

\subsection{Implementation of Efficient Baselines.}
\label{app:efficient_baselines}
To ensure a fair comparison, we re-implemented representative efficient retrieval methods within our decoder-only framework.
For KALE\citep{campos2023quick} and \citet{wang2023query}, we initialized the query encoder using the first 3 layers of Qwen3-4B (approximately 303M parameters excluding the LM head) with an embedding dimension of 2560.
For ScalingNote\citep{huang2024scalingnote}, we employed Medical-Embedder-base as the query encoder.
For the Distill baseline~\citep{ren2021rocketqav2}, following standard protocols, we utilized scores from Medical-Embedder-4B as soft labels and optimized the student model (Medical-Embedder-base) using a combination of KL-divergence and InfoNCE loss.

\subsection{Training Details}
\label{app:hyperparams}
For the query encoder, we use the final hidden state of the [CLS] token as the sentence embedding. For the document encoder, we append an [EOS] token to the input sequence and use its output hidden state as the document embedding.
The maximum input length for both queries and documents is set to 512 tokens. 

We summarize the training configurations in Table~\ref{tab:full_hyperparams}.  
For memory efficiency, we enable gradient checkpointing and use DeepSpeed Stage 0. For models up to 4B parameters, we train in \texttt{fp16}, while for the 8B model we switch to \texttt{bf16} to ensure stability.  
All document encoders are fine-tuned with LoRA (rank $32$, $\alpha=64$). 
For fair comparison, all symmetric baseline models are fine-tuned for 2 epochs, matching the total exposure of our asymmetric models, which observe the fine-tuning data once during independent initialization and once again during joint fine-tuning. 

In our asymmetric architecture, both query and document encoders are first initialized by one epoch of fine-tuning.  
For Stage~\uppercase\expandafter{\romannumeral1}, we align query and document embeddings using $8.4$M pairs of query alignment data for one epoch.  
For Stage~\uppercase\expandafter{\romannumeral2}, we further fine-tune for one epoch to ensure comparability with other baselines.  
We apply the same learning rate ($1\times10^{-4}$) to both query and document encoders, as we observed that asymmetric learning rates led to performance degradation.  

\begin{table*}[t]
    \centering
    
    \resizebox{0.8\textwidth}{!}{ 
    \begin{tabular}{l ccc c cc}
        \toprule
        \multirow{2}{*}{\textbf{Hyperparameter}} & 
        \multicolumn{3}{c}{\textbf{General Training}} & & 
        \multicolumn{2}{c}{\textbf{Asymmetric Training}} \\
        
        \cmidrule(lr){2-4} \cmidrule(lr){6-7}
        
         & \textbf{RetroMAE} & \textbf{Unsup.} & \textbf{Fine-tuning} & & 
         \textbf{Stage \uppercase\expandafter{\romannumeral1}} & 
         \textbf{Stage \uppercase\expandafter{\romannumeral2}} \\
        \midrule
        
        Peak Learning Rate & 2e-4 & 1e-4 & 1e-4 & & 1e-4 & 1e-4 \\
        Warmup Ratio       & 0.0  & 0.05 & 0.05 & & 0.05 & 0.05 \\
        Global Batch Size  & 384  & 19,200 & 640 & & 2,560 & 640  \\
        Epochs             & 3    & 3      & 2   & & 1     & 1    \\
        \bottomrule
    \end{tabular}
    }
    \caption{
        \textbf{Detailed Training Hyperparameters.} 
        We list the configurations for both the general pre-training stages (RetroMAE, Unsupervised) and our proposed asymmetric alignment stages (Stage \uppercase\expandafter{\romannumeral1} \& \uppercase\expandafter{\romannumeral2}). 
        All stages utilize the AdamW optimizer with a linear decay scheduler.
    }
    \label{tab:full_hyperparams}
\end{table*}

\section{Cost-Benefit Analysis}
\label{app:cost-benefit}

We address the cost-benefit concern by providing detailed offline metrics. In industrial retrieval systems, document indices are updated much less frequently than queries are processed. Therefore, our architecture strategically leverages powerful offline models to boost representation quality without penalizing the critical online user experience.

\paragraph{Offline Efficiency.} As shown in Table~\ref{tab:offline_efficiency}, processing 1 million documents with our largest 8B model takes only 3.77 hours on a single A100 GPU. Given the low update frequency (monthly or weekly) typical in production medical retrieval systems, this one-time offline cost is negligible compared to the continuous benefits in online retrieval accuracy. Although \texttt{gte-multilingual-base} is faster (0.21h), the 8B model's overhead is entirely acceptable for periodic updates. Note that index memory is constant as MRL aligns the document embedding dimension with the query encoder regardless of model size.

\begin{table}[h]
\centering
\small
\setlength{\tabcolsep}{4pt}
\begin{tabular}{lccc}
\toprule
\textbf{Model} & \textbf{Offline Params} &  \textbf{Time (h)} \\
\midrule
gte-multilingual-base & 305M  & 0.21 \\
CARE-0.3B-4B & 4B  & 2.95 \\
CARE-0.3B-8B & 8B  & 3.77 \\
\bottomrule
\end{tabular}
\caption{Offline Efficiency Analysis. Encoding time for 1M documents on A100 80GB. Index memory constant (2,929.69 MB) due to MRL dimension alignment.}
\label{tab:offline_efficiency}
\end{table}

\paragraph{Online Advantage.} Crucially, the primary bottleneck in real-world applications lies in online inference latency rather than offline costs. While offline costs increase marginally, CARE-0.3B-8B achieves a superior average score of 78.94 with only 0.3B online parameters. It surpasses the fine-tuned \texttt{gte-Qwen2-1.5B} by +1.33 with 9$\times$ higher QPS. This decoupled design allows independent scaling: document encoders can be upgraded for accuracy gains without any impact on online serving latency.

\paragraph{Training Cost.} We present a direct comparison of GPU hours in Table~\ref{tab:training_cost}. Our two-stage training strategy is highly efficient. For example, Stage 2 for the 4B model requires 260 hours, comparable to standard fine-tuning of the document encoder alone (280 hours). While Stage 1 represents an initial alignment cost, it establishes a reusable foundation. This confirms that our approach improves performance without introducing prohibitive computational costs compared to conventional fine-tuning paradigms.

\begin{table}[t]
\centering
\small
\resizebox{0.5\textwidth}{!}{
\begin{tabular}{lccc}
\toprule
\textbf{Model} & \textbf{Stage-1 (h)} & \textbf{Stage-2 (h)} & \textbf{Finetune Baseline (h)} \\
\midrule
CARE-0.3B-4B & 316 & 260 & 280 \\
CARE-0.3B-8B & 608 & 529 & 544 \\
\bottomrule
\end{tabular}
}
\caption{Training Cost Analysis (GPU Hours on A100-40G).}
\label{tab:training_cost}
\end{table}

\section{Quality Analysis of Open-Source Benchmarks}
\label{app:open_source_analysis}
We investigate the false negative issue in {CmedqaRetrieval} and {MedicalRetrieval}. 
For each query, we use {gte-multilingual-base} to retrieve the top-50 candidate documents and re-annotate them using GPT-4o under a 5-point relevance scale with prompt in Table~\ref{tab:prompt_medteb1}. 

Results in Table~\ref{tab:llm_reannotation} suggest that a large number of retrieved documents, though unlabeled in the original datasets, are judged as relevant by the LLM. Table~\ref{tab:false_neg_cmedqa} and Table~\ref{tab:false_neg_medretri} shows several examples of false negatives and false positives, together with the topic intensity phenomenon in medical domain that certain diseases or drugs generate a large volume of semantically similar queries and answers.
This indicates annotation incompleteness in existing benchmarks. 
These findings raise concerns about the validity of current benchmarks for a reliable evaluation of medical retrieval capability.

\begin{table*}[h]
\centering
\resizebox{0.7\textwidth}{!}{
\begin{tabular}{lrrrr}
\toprule
\textbf{Benchmark} & \textbf{Orig. Pos.}& \textbf{LLM-Labeled Pos.}& \textbf{False Positive}& \textbf{False Negative}\\
\midrule
MedicalRetrieval & 0.81& 9.11 & 0.26 & \textbf{8.56}\\
CmedqaRetrieval  & 1.42 & 19.94 & 0.46 & \textbf{18.98} \\
\bottomrule
\end{tabular}
}
\caption{LLM re-annotation on open-source medical retrieval benchmarks. To aid interpretation, we assume the LLM labels are pseudo–ground truth.  
We measure the average number of positive documents per query in the original dataset vs. LLM-labeled data, and identify false positives and false negative.}
\label{tab:llm_reannotation}
\end{table*}

\begin{table*}[t]
    \centering
    \scalebox{1.00}{
    \begin{small}
    \begin{tabular}{p{0.95\linewidth} }
        \toprule
        \textbf{Query} \\
        \begin{CJK}{UTF8}{gkai} 查出说是贫血孩子老烧还有咳嗽
        \end{CJK} \\
        \textit{The child was diagnosed with anemia and has been running a fever with coughing.}  \\
        \midrule
        \textbf{False Positive (Labeled as positive, but not mention fever and coughing.)} \\
        \begin{CJK}{UTF8}{gkai}如果是检查有贫血，可以结合贫血的类型和严重的程度，根据检查结果进行治疗的考虑即可。
        \end{CJK} \\
        \textit{If anemia is detected, treatment can be determined based on the type and severity of anemia, as indicated by the test results.} \\
        \midrule
        \textbf{False Negative (Labeled as negative, but annotated as positive by LLM)} \\
        \begin{CJK}{UTF8}{gkai}如果只是简单地烧咳嗽，等相应症状，是无需，担心的，但是贫血的原因必须地须要查清楚，一般情况下评选分为营养不良性贫血，还有其他病理性贫血。所以建议到医院进行系统检查，看到底是？出现的什么方面的贫血。然后对症治疗。
        \end{CJK} \\
        \textit{If it is just a simple cough or other corresponding symptoms, there is no need to worry. However, the cause of anemia must be clarified. Generally, anemia is categorized into nutritional deficiency anemia and other pathological anemias. Therefore, it is recommended to go to the hospital for a comprehensive examination to determine the specific type of anemia and then treat it accordingly.} \\
        \bottomrule
    \end{tabular}
    \end{small}
    }
    \caption{
    An example of false negatives in CmedqaRetrieval. 
    }
    \label{tab:false_neg_cmedqa}
\end{table*}

\begin{table*}[t]
    \centering
    \scalebox{1.00}{
    \begin{small}
    \begin{tabular}{p{0.95\linewidth} }
        \toprule
        \textbf{Query} 
        \\
        \begin{CJK}{UTF8}{gkai} 感冒发烧一起来怎么办
        \end{CJK} \\
        \textit{What should I do if I have a cold and fever at the same time?} \\
        \midrule
        \textbf{Positive} 
        \\ 
        \begin{CJK}{UTF8}{gkai}	你好，应该是流行感冒吧，典型流感，急起高热，全身疼痛，显著乏力，呼吸道症状较轻。颜面潮红，眼结膜外眦充血，咽充血，软腭上有滤泡。具体吃药建议咨询医生。 期间注意：多饮开水，多食新鲜的蔬菜、瓜果。饮食宜清淡，多食易消化，且富于营养及富含维生素的食物，如稀饭、豆浆、面条、水果等。窗子经常开下来通通风。 一些简单食疗： 1.生姜20克、大蒜头5～6瓣、红糖适量。 用法：水煎服。 主治：流行性感冒初起，头痛，怕冷发热，无汗，伴有恶心者。 说明：流行性感冒是一种急性呼吸道传染病，表现为急起高热，全身疼痛，乏力，呼吸道症状如咽干喉痒，干咳等，胃肠道症状如恶心呕吐、腹泻水样便等。 2.冬瓜粥 粳米50克。将冬瓜适量切成小块，与米同煮，粥熟即可食用。此粥对病毒型流行性感冒病人有效。 3.葱白500克、大蒜250克。 用法：上药切碎加水2000毫升煎煮。日服3次，每次250毫升，连服2～3天。 愿早日康复！
        \end{CJK} \\
        \textit{Hello, it sounds like you have the flu. Typical symptoms include sudden high fever, body aches, significant fatigue, and mild respiratory symptoms. You may also have facial flushing, conjunctival injection, pharyngeal congestion, and follicles on the soft palate. Please consult a doctor for specific medication advice. During this time, drink plenty of water and eat more fresh vegetables and fruits. Keep your diet light and easy to digest, focusing on nutritious and vitamin-rich foods like porridge, soy milk, noodles, and fruits. Ventilate your room regularly by opening windows. Here are some simple home remedies: 1. 20 grams of fresh ginger, 5–6 cloves of garlic, and an appropriate amount of brown sugar. Decoct in water and take orally. This is for the early stages of influenza with headache, chills, fever, no sweating, and nausea. Influenza is an acute respiratory infectious disease characterized by sudden high fever, body aches, fatigue, and respiratory symptoms like sore throat and dry cough. It may also cause gastrointestinal symptoms like nausea, vomiting, and watery diarrhea. 2. Winter melon porridge: 50 grams of japonica rice. Cut an appropriate amount of winter melon into small pieces and cook with rice. This porridge is effective for patients with viral influenza. 3. 500 grams of green onion whites and 250 grams of garlic. Chop the ingredients and decoct in 2000 milliliters of water. Take three times a day, 250 milliliters each time, for 2–3 days. Hope you recover soon!} \\

        \midrule

        \textbf{False Negative 1 (Labeled as negative, but annotated as positive by LLM)} 
        \\ 
        \begin{CJK}{UTF8}{gkai}建议口服抗病毒药物和感冒冲剂试试.有炎症还是应该加上抗生素。口服药物不见效的，建议输液治疗为好。在当地医生指导下使用。发烧用退热贴
        \end{CJK} \\
        \textit{It is suggested to try oral antiviral medications and cold granules. If there is an infection, antibiotics should be added. If oral medications are not effective, it is recommended to consider intravenous therapy. This should be done under the guidance of a local doctor. For fever, you can use fever patches.} \\
        \textbf{False Negative 2} \\
        \begin{CJK}{UTF8}{gkai}感冒发烧是临床上最常见的疾病和症状，具体吃药要根据具体的症表现以及病人身体状况而定。如果是儿童出现感冒发烧的情况一般选择以单药为主，出现发烧时主要可选择对乙酰氨基酚或者布洛芬口服液来进行治疗；如果还有其他的症状，比如出现鼻塞流涕，可以使用氨咖黄敏颗粒。如果是成人感冒发烧，一般多选择复合剂型，比如酚麻美敏片或者复方氨酚烷胺等。如果持续发烧不退，要及时完善血液分析和胸片检查排除并发肺炎的可能。
        \end{CJK} \\
        \textit{A cold with fever is one of the most common illnesses and symptoms clinically. The specific medication should be determined based on the specific symptoms and the patient's physical condition. For children with a cold and fever, monotherapy is usually chosen. For fever, acetaminophen or ibuprofen oral suspension can be used for treatment. If there are other symptoms, such as nasal congestion and runny nose, pheniramine and caffeine granules can be used. For adults with a cold and fever, compound formulations are generally preferred, such as phenylephrine, dextromethorphan, and acetaminophen tablets, or compound paracetamol and amantadine. If the fever persists, it is important to promptly complete blood tests and chest X-rays to rule out the possibility of pneumonia.} \\
        \textbf{False Negative 3} \\
        \begin{CJK}{UTF8}{gkai}你好，建议口服抗病毒药物和感冒冲剂试试.即使是病毒性感冒也容易继发细菌感染,所以最好还是应该加上抗生素口服.建议口服药物不见效的,建议输液抗炎治疗为好.因为还是输液血药浓度更高见效更快更好啊.有痰的加上鲜sd竹沥口服试试.发烧还需要适当加上额外的退烧药物.一般需要7-10天才能治愈的.最好还是看医生啊
        \end{CJK} \\
        \textit{Hello, it is suggested to try oral antiviral medications and cold granules. Even viral colds can easily lead to secondary bacterial infections, so it is better to add oral antibiotics. If oral medications are not effective, it is recommended to consider intravenous anti-inflammatory treatment, as it provides higher blood drug concentration and faster results. For those with phlegm, you can try adding fresh bamboo extract orally. Fever also requires the addition of extra antipyretic drugs. It usually takes 7-10 days to recover. It is best to see a doctor.} \\
        \bottomrule
    \end{tabular}
    \end{small}
    }
    \caption{
    An example of false negatives in MedicalRetrieval. This example shows the \textit{topic intensity} phenomenon in medical domain: certain diseases or drugs generate a large volume of semantically
similar queries and answers.
    }
    \label{tab:false_neg_medretri}
\end{table*}

\section{Annotation Prompts}
\label{app:prompts}
Table~\ref{tab:prompt_medteb1} and Table~\ref{tab:prompt_medteb2} present the prompts templates for CMedTEB construction. Table~\ref{tab:train_prompt} shows prompt for our training data annotation.

\begin{table*}[htbp]
\centering
\small
\begin{tabular}{l}
\hline
\toprule
Prompt: \\
This is a medical information retrieval task: given a medical query (Query), retrieve \\documents (Passages)that can answer the question. \\
\\
Given a medical query (Query) and \{len(docs)\} passages, your task is to rate the \\relevance between the Query and each Passage. \\
\\
Relevance scoring criteria: \\
S: The subject (e.g., disease name, drug name, inquiry target) and inten of Query and \\Passage are fully consistent.  The Passage can directly, completely, and correctly  answer the Query. \\
A: The subject and intent of Query and Passage are consistent. The Passage contains \\content that candirectly and correctly answer the Query. \\
B: The subject of Query and Passage is consistent, but the intent differs. \\The Passage cannot directly answer the Query, but it is useful for inference. \\
C: The subject of Query and Passage is related, but the intent is inconsistent. \\It can only partially match the Query from the text, but  cannot answer the Query. \\
D: The subject and intent of Query and Passage are unrelated. Cannot answer the Query. \\
\\
Notes: \\
1. Query and Passage are independent; there is no contextual relationship.\\ Do not infer or supplement the subject/intent of Query based on Passage. \\
2. If the Query is low-quality (e.g., missing subject, like "How to treat this disease?"),\\ the maximum relevance score for all Passages should not exceed B. \\
3. All Passages are independent; they are randomly ordered and have no contextual relationship. \\
\\
Output format: \\
Your output must be a JSON object, containing only the required fields. The format is as follows: \\
\{
  "Passage-0": "A",
  "Passage-1": "C",
  ...
\}
\\
Query and Passages are as follows: \\
- Query: \{query\} \\
\{passages\} \\
... \\
Remember: do not output any other content or explanation. \\Your output must be only a JSON object with the required fields. Output:\\
\bottomrule
\hline
\end{tabular}
\caption{Prompt template for CMedTEB Retrieval and Rerank tasks}
\label{tab:prompt_medteb1}
\end{table*}

\begin{table*}[htbp]
\centering
\small
\begin{tabular}{p{0.95\linewidth}}
\hline
Medical Query Rewriting Sample Generation (Positive and Negative Examples) \\ \\
Task Objective: Your task is to generate one positive example and two negative examples based on a given original medical query and a set of synonyms. \\
You will receive a JSON object containing the following fields: \\
"origin": "Original medical term", \\
"replace": "Synonym medical term for replacement", \\
"query\_pairs": \{ "origin": "Query sentence using the original term", "replace": "Query sentence using the replaced term" \} \\

\\
Generation Rules: \\
1. General Quality Standards (applicable to all outputs): \\
- Professional Expression: Use professional, fluent, and natural medical language. \\
- Medical Accuracy: Content must conform to medical knowledge and avoid ambiguity. \\
- Format Requirement: All outputs must be complete, fluent interrogative sentences. \\

\\
2. Specific Sample Requirements: \\
- positive (Positive Example): \\
  - Task: Optimize and rewrite the second query in query\_pairs (the one containing the "replace" term). \\
  - Intent: Must preserve the exact same intent as the original query. \\
  - Terminology: Must use the term specified in the "replace" field. \\
  - Constraint: Rewritten query length must be within ±30\% of the original query’s length. \\

- negative-1 (Negative Example 1): \\
  - Task: Create a new query based on the topic of the original query, similar but distinctly different. \\
  - Terminology: Must use the term specified in the "replace" field. \\
  - Intent: Significantly alter the intent of the original query. \\

- negative-2 (Negative Example 2): \\
  - Task: Create a new query based on the topic of the original query, similar but distinctly different. \\
  - Terminology: Must use the term specified in the "origin" field. \\
  - Intent: Significantly alter the intent of the original query (same rule as negative-1). \\

\\
Output Format: Must be a JSON object containing only the following three fields. Do not add any extra explanations or comments. \\

\\
Input: \{input\}\\

Output: \\
\hline
\end{tabular}
\caption{Prompt template for CMedTEB STS tasks}

\label{tab:prompt_medteb2}
\end{table*}

\begin{table*}[htbp]
\centering
\small
\begin{tabular}{p{0.95\linewidth}}
\hline
This is a retrieval task in the Chinese medical domain, requiring classification of positive and negative documents based on the user’s medical query and search engine returned documents.\\

You will receive data containing the following fields: \\
"query": User input in the medical domain. \\
"documents": A candidate document set containing multiple documents, some relevant and some irrelevant — capable or incapable of answering the query. \\

Your task is to identify "positive\_document" and "negative\_document" from the provided documents. \\
"positive\_document": Relevant to the query; the document contains sentences that can answer the query. \\
"negative\_document": Either relevant or irrelevant to the query, but the document content does NOT contain any sentence that can answer the query. \\

Please follow these guidelines: \\
- Both "positive\_document" and "negative\_document" must come from the candidate document set. \\
- "positive\_document" and "negative\_document" are mutually exclusive — no document overlap is allowed. \\

Output Requirements: \\
Example: \{out\_exam\}. \\
Your output must always be ONLY a JSON object, containing ONLY document indices (e.g., "doc-1"). Do NOT include document content, explanations, or any additional text. \\

Input Data Format: \\
\{"positive\_document":["doc-1","doc-2"], "negative\_document":["doc-3","doc-4"]\} \\
Classify the documents in the input data according to the above rules, ensuring the output strictly follows the required format. \\
Output: \\
\hline
\end{tabular}
\caption{Prompt template for training data annotation}
\label{tab:train_prompt}
\end{table*}


\end{document}